\documentclass[acmtog, nonacm]{acmart}

\usepackage{geometry}
\usepackage{multirow}
\usepackage{colortbl}
\usepackage{pifont}
\usepackage{booktabs}
\usepackage{graphicx}
\usepackage[table]{xcolor}
\usepackage{subcaption}
\usepackage{amsmath}

\definecolor{colorfirst}{HTML}{ffb2b3}  
\definecolor{colorsecond}{HTML}{ffd9b6} 
\definecolor{colorthird}{HTML}{ffffb9}  

\newif\ifrevisionmarks
\definecolor{revisionblue}{RGB}{0,0,255}
\definecolor{revisionorange}{RGB}{230,126,34}
\newenvironment{addedblock}{\ifrevisionmarks\color{revisionblue}\fi}{}
\newenvironment{revisedblock}{\ifrevisionmarks\color{revisionorange}\fi}{}
\newcommand{\added}[1]{{\ifrevisionmarks\color{revisionblue}\fi #1}}
\newcommand{\revised}[1]{{\ifrevisionmarks\color{revisionorange}\fi #1}}
\AtBeginDocument{%
  }

\setcopyright{acmlicensed}
\copyrightyear{2026}
\acmYear{2026}
\acmDOI{XXXXXXX.XXXXXXX}

\acmJournal{TOG}

\begin{document}

\title{ExMesh++: From Multi-View Images to Relightable UV-PBR Mesh Assets
via Topology-Adaptive Reconstruction and Decomposition}

\author{Chuanjin Fan}
\affiliation{%
  \institution{University of Science and Technology of China}
  \city{Hefei}
  \country{China}}
\email{fancj@mail.ustc.edu.cn}

\author{Lifan Wu}
\affiliation{%
  \institution{University of Science and Technology of China}
  \city{Hefei}
  \country{China}}
\email{wusar@mail.ustc.edu.cn}

\author{Wenjie Chang}
\affiliation{%
  \institution{Alibaba Group, Amap}
  \city{Beijing}
  \country{China}}
\email{changwj@mail.ustc.edu.cn}

\author{Hanzhi Chang}
\affiliation{%
  \institution{University of Science and Technology of China}
  \city{Hefei}
  \country{China}}
\email{changhz@mail.ustc.edu.cn}

\author{Wenfei Yang}
\affiliation{%
  \institution{University of Science and Technology of China}
  \city{Hefei}
  \country{China}}
\email{yangwf@ustc.edu.cn}

\author{Tianzhu Zhang}
\affiliation{%
  \institution{University of Science and Technology of China}
  \city{Hefei}
  \country{China}}
\email{tzzhang@ustc.edu.cn}

\renewcommand{\shortauthors}{Fan et al.}

\begin{abstract}
\begin{addedblock}
Multi-view reconstruction extends beyond surface recovery to editable and relightable mesh assets. Such assets require well-formed topology, valid UV parameterization, and explicit PBR material maps. Existing surface reconstruction approaches optimize implicit fields, Gaussian primitives, or other intermediate representations. Converting them into such assets often requires surface extraction and texture baking. Inverse-rendering methods estimate materials and illumination, yet these components often remain tied to neural fields or point-based primitives rather than the final mesh. Joint optimization of geometry, materials, and lighting may also allow these variables to compensate for one another, leading to ambiguous decomposition. To address these limitations, we present ExMesh++, a staged framework for reconstructing relightable UV-PBR mesh assets from multi-view images. The first stage refines explicit mesh geometry and topology through adaptive vertex splitting and merging, while maintaining UV consistency as the topology changes. The second stage fixes the resulting mesh-UV carrier and optimizes UV-space PBR maps together with environment lighting. Building on this stable carrier, ExMesh++ models one-bounce diffuse indirect illumination through secondary-ray tracing with shared UV-PBR materials. Experiments demonstrate competitive geometry accuracy, strong relighting performance, and direct usability of the exported assets in standard DCC workflows.
\end{addedblock}
\end{abstract}




\begin{teaserfigure}
  \Description{teaser.}
  \centering
  \includegraphics[width=1.0\textwidth]{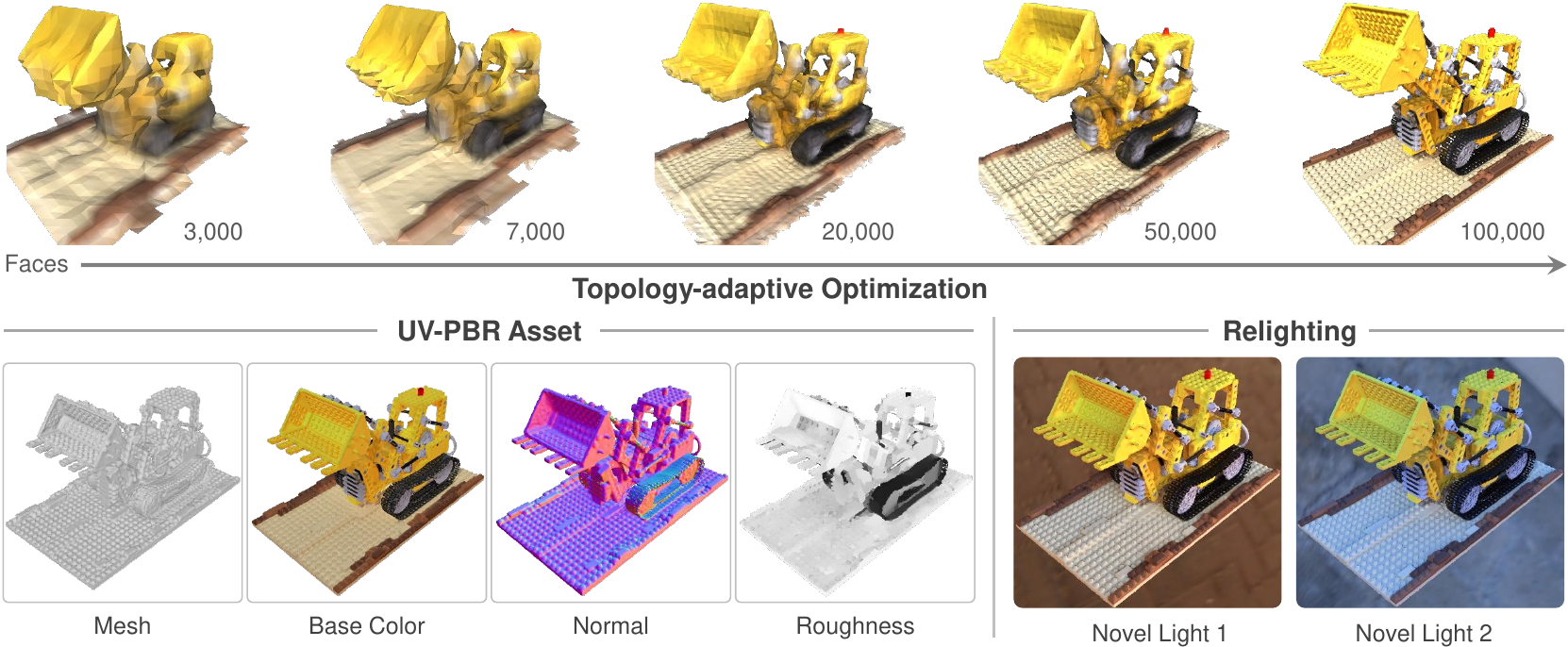}
  \caption{\added{Overview of the reconstruction and asset generation results produced by ExMesh++. The top row shows the topology-adaptive optimization process, where a coarse mesh is progressively refined into a compact high-resolution surface. The bottom row shows the exported UV-PBR asset, including the mesh, base-color, normal, and roughness maps, together with relighting results under novel environment illumination.}}
  \label{fig:teaser}
\end{teaserfigure}

\maketitle
\begin{addedblock}
\section{Introduction}
\label{sec:introduction}

Reconstructing 3D objects from multi-view images has long been a central problem in connecting 2D observations with editable 3D representations. Traditional surface reconstruction mainly focuses on geometric accuracy, surface completeness, or novel-view rendering quality, while practical graphics workflows require assets that can be edited and relit in standard rendering pipelines. This raises a stronger requirement for reconstruction methods: the output should not be merely an accurate mesh, but a mesh-UV-PBR asset.

Recent years have witnessed rapid progress in surface reconstruction, from neural implicit representations to explicit point-based primitives. NeRF-based methods~\cite{mildenhall2020nerf} synthesize high-quality novel views through volume rendering, while NeuS~\cite{wang2021neus}, VolSDF~\cite{yariv2021volsdf}, and Neuralangelo~\cite{tancik2023neuralangelo} further advance neural surface reconstruction. Later, 3D Gaussian Splatting~\cite{kerbl2023gaussian} greatly improved training and rendering efficiency, inspiring a series of surface-oriented variants such as 2DGS~\cite{huang2024twodgs}, GOF~\cite{yu2024gof}, PGSR~\cite{chen2024pgsr}, and QGS~\cite{zhang2025qgs}. In parallel, several mesh-driven methods attempt to reduce the gap between intermediate representations and meshes by using differentiable tetrahedral grids~\cite{shen2021dmtet,munkberg2022nvdiffrec}, point clouds~\cite{yang2025imlssplatting}, or sparse voxel representations~\cite{li2025geosvr}. Despite their different optimization representations, these methods still obtain the final mesh through surface extraction~\cite{lorensen1987marchingcubes} from implicit fields, Gaussian primitives, or other intermediate representations. Their appearance representations are also usually designed for fitting the original observations, such as baked RGB textures or view-dependent radiance. Although such outputs are effective for reconstructing the captured views, they lack an explicit decomposition of materials and illumination. As a result, they are difficult to directly use for physically based re-rendering or for material-level editing.

A natural solution is to introduce inverse rendering. Instead of only fitting image appearance, inverse rendering decomposes multi-view observations into geometry, materials, and illumination. This allows the reconstructed representation to be rendered under novel lighting. Early representative methods model relightable scenes with neural fields or tensor factorization~\cite{zhang2022invrender,jin2023tensoir}. Recent works further extend this idea to Gaussian representations, enabling more efficient relightable reconstruction~\cite{liang2024gsir,gao2024relightable3dgs,gu2025irgs}. However, their optimization carriers are still neural fields or point-based primitives. Although these methods can produce relightable results, the optimized representation is still separated from the final mesh-UV-PBR asset required by production pipelines. Additional surface extraction, material baking, or format conversion is often required to turn the result into an editable asset for standard rendering tools.

To recover mesh and UV-PBR material assets from multi-view images, NVDiffRecMC~\cite{hasselgren2022nvdiffrecmc} places geometry, spatially varying materials, and environment lighting into a unified inverse rendering framework. It uses Monte Carlo rendering~\cite{kajiya1986rendering} to improve the physical consistency of material-light decomposition. This line of work shows that recovering relightable assets from images is feasible. However, for general object asset reconstruction, the final asset should be built on a stable mesh-UV carrier. If geometry, normal variation, albedo, roughness, and illumination are all adjusted throughout the whole optimization process, photometric errors may be absorbed by multiple factors at the same time. This ambiguity can entangle geometric reconstruction with material and lighting estimation. Such a tightly coupled optimization is effective for image fitting, but may make both geometric reconstruction and material decomposition less reliable.

Based on this observation, we extend our prior ExMesh framework \cite{Fan2026ExMesh} into ExMesh++, a staged framework for explicit asset reconstruction. The first stage constructs the asset carrier. It directly optimizes the geometry and topology of meshes through adaptive vertex splitting and merging, while maintaining the UV parameterization during topology changes. This stage produces compact and clean meshes with a stable UV atlas and an RGB appearance texture. The second stage freezes the vertex positions, topology, and UV coordinates. It then optimizes PBR material maps and environment lighting in the shared UV space, including base color, roughness, normal maps, an optional metallic map, and an environment map. By fixing the mesh-UV carrier before PBR optimization, ExMesh++ reduces compensation among geometry, materials, and lighting, while keeping the optimized maps attached to the final exported asset. The fixed mesh-UV carrier also enables one-bounce diffuse indirect illumination. We trace secondary rays on the reconstructed mesh and query the same PBR maps to capture local color bleeding, without introducing an additional learned residual appearance field.

To verify both the reconstructed geometry and the decomposed asset, we evaluate ExMesh++ across both reconstruction quality and asset usability. On the DTU~\cite{jensen2014dtu} and Stanford-ORB~\cite{kuang2023stanfordorb} datasets, we evaluate the geometric accuracy and structural quality of meshes. On the Synthetic4Relight~\cite{zhang2022invrender} and Stanford-ORB datasets, we evaluate novel view synthesis under the original illumination and relighting under multiple environment maps. On DTU, we further show the effect of one-bounce indirect illumination on local color bleeding and occluded regions. Finally, we place the exported mesh, PBR maps, and environment lighting into a standard DCC workflow to demonstrate relighting, texture editing, and scene composition with artist-created assets. Experiments show that ExMesh++ achieves competitive geometry, NVS, and relighting results, while producing explicit UV-PBR assets that can be directly used in downstream workflows.

Our main contributions are summarized as follows:

\begin{enumerate}
	\item \begin{revisedblock} We develop a topology-adaptive explicit mesh reconstruction framework that directly optimizes mesh geometry and topology through vertex splitting and merging, with consistent UV updates during topology changes.\end{revisedblock}
	\item We formulate UV-PBR inverse rendering as a second-stage optimization on the reconstructed mesh-UV carrier, reducing geometry-appearance compensation by freezing geometry during material-light decomposition.
	\item We incorporate one-bounce diffuse indirect illumination by tracing secondary rays on the fixed mesh and querying the shared UV-PBR materials, without introducing an additional learned residual appearance field.
	\item Experiments demonstrate both reconstruction quality and asset usability, showing competitive results and downstream applications in relighting and texture editing.
\end{enumerate}

\end{addedblock}
\begin{revisedblock}
\section{Related Work}
\label{sec:related-work}

\subsection{Multi-View Surface Reconstruction}

\textbf{Neural implicit surface reconstruction.} In recent years, neural implicit representations have become a central approach to multi-view surface reconstruction. NeRF~\cite{mildenhall2020nerf} learns a continuous radiance field through volume rendering. To recover more clearly defined surfaces, IDR~\cite{yariv2020multiview}, UNISURF~\cite{oechsle2021unisurf}, NeuS~\cite{wang2021neus}, and VolSDF~\cite{yariv2021volsdf} represent geometry with occupancy or signed distance fields and optimize them through differentiable rendering. Geo-NeuS~\cite{fu2022geoneus} and Neuralangelo~\cite{tancik2023neuralangelo} further improve reconstruction accuracy and detail by introducing multi-view geometric priors and multi-resolution encodings, respectively. However, the optimized representation remains an implicit field, and a mesh must be extracted afterward, typically with Marching Cubes~\cite{lorensen1987marchingcubes}. This limits direct control over mesh connectivity and quality. The extraction step may also smooth sharp features or fail to preserve thin structures.

\textbf{Gaussian-based surface reconstruction.} 3D Gaussian Splatting~\cite{kerbl2023gaussian} provides efficient optimization and real-time rendering, but its unstructured volumetric primitives do not directly define a surface. Subsequent work improves surface awareness through primitive design or mesh extraction. SuGaR~\cite{su2023sugar} regularizes Gaussians toward local surfaces. 2DGS~\cite{huang2024twodgs} uses oriented disks, while PGSR~\cite{chen2024pgsr} and QGS~\cite{zhang2025qgs} adopt planar and quadric primitives, respectively. GOF~\cite{yu2024gof} instead builds a continuous opacity field and extracts its level set. Although these approaches improve local surface modeling, the final mesh is still produced through TSDF fusion~\cite{curless1996volumetric, newcombe2011kinectfusion}, Poisson reconstruction~\cite{kazhdan2006poisson}, or isosurface extraction~\cite{lorensen1987marchingcubes}. As a result, mesh connectivity and compactness are not direct optimization objectives.

\subsection{Differentiable Mesh Optimization}

\textbf{Geometry and topology optimization.} Differentiable rasterizers such as Neural Mesh Renderer~\cite{kato2018neuralmeshrenderer}, Soft Rasterizer~\cite{liu2019softrasterizer}, and nvdiffrast~\cite{laine2020modular} allow image-space losses to propagate directly to mesh vertices, forming the basis of explicit mesh optimization. Template-based approaches directly deform a predefined mesh~\cite{wang2018pixel2mesh}. DMTet~\cite{shen2021dmtet}, NVDiffRec~\cite{munkberg2022nvdiffrec}, and FlexiCubes~\cite{shen2023flexicubes} support topology changes through tetrahedral grids or scalar fields, but the final mesh is still extracted from an intermediate representation. DMesh~\cite{son2024dmesh} and DMesh++~\cite{son2025dmeshpp} instead formulate mesh geometry and connectivity within differentiable mesh representations. In general, fixed connectivity limits adaptive refinement, while dynamic connectivity may introduce irregular triangles, degenerate faces, or redundant elements. This creates a need to balance local refinement with structural simplification.

\textbf{Texture representations and UV parameterization.} Mesh appearance is commonly represented by per-vertex colors, per-face attributes, or features attached to geometric primitives. These representations couple appearance resolution with geometric density and may require denser meshes for high-frequency detail. Neural texture methods, including Texture Fields~\cite{oechsle2019texturefields}, NeuTex~\cite{xiang2021neutex}, NVDiffRec~\cite{munkberg2022nvdiffrec}, and NeuMesh~\cite{yang2022neumesh}, use continuous functions, learned mappings, or latent features to reduce this dependence. However, converting these representations into standard texture maps often requires additional baking or format conversion.

Traditional parameterization methods such as LSCM~\cite{levy2002lscm} and ABF++~\cite{sheffer2005abfpp} generate UV atlases for fixed meshes. Joint optimization of UV mappings and texture baking has also been explored for fixed meshes~\cite{10.1145/3617683}. These methods generally assume unchanged connectivity. When vertices and faces are inserted or removed during optimization, face-corner UV indices, seams, and UV islands must be updated accordingly. Reparameterization may also disrupt the correspondence between the optimized texture and the surface. Maintaining a valid UV parameterization during topology changes therefore remains challenging.

\begin{figure*}[t]
  \Description{Pipeline overview of ExMesh++ from multi-view images to a mesh-UV carrier and UV-PBR asset decomposition.}
  \centering
  \includegraphics[width=1.0\textwidth]{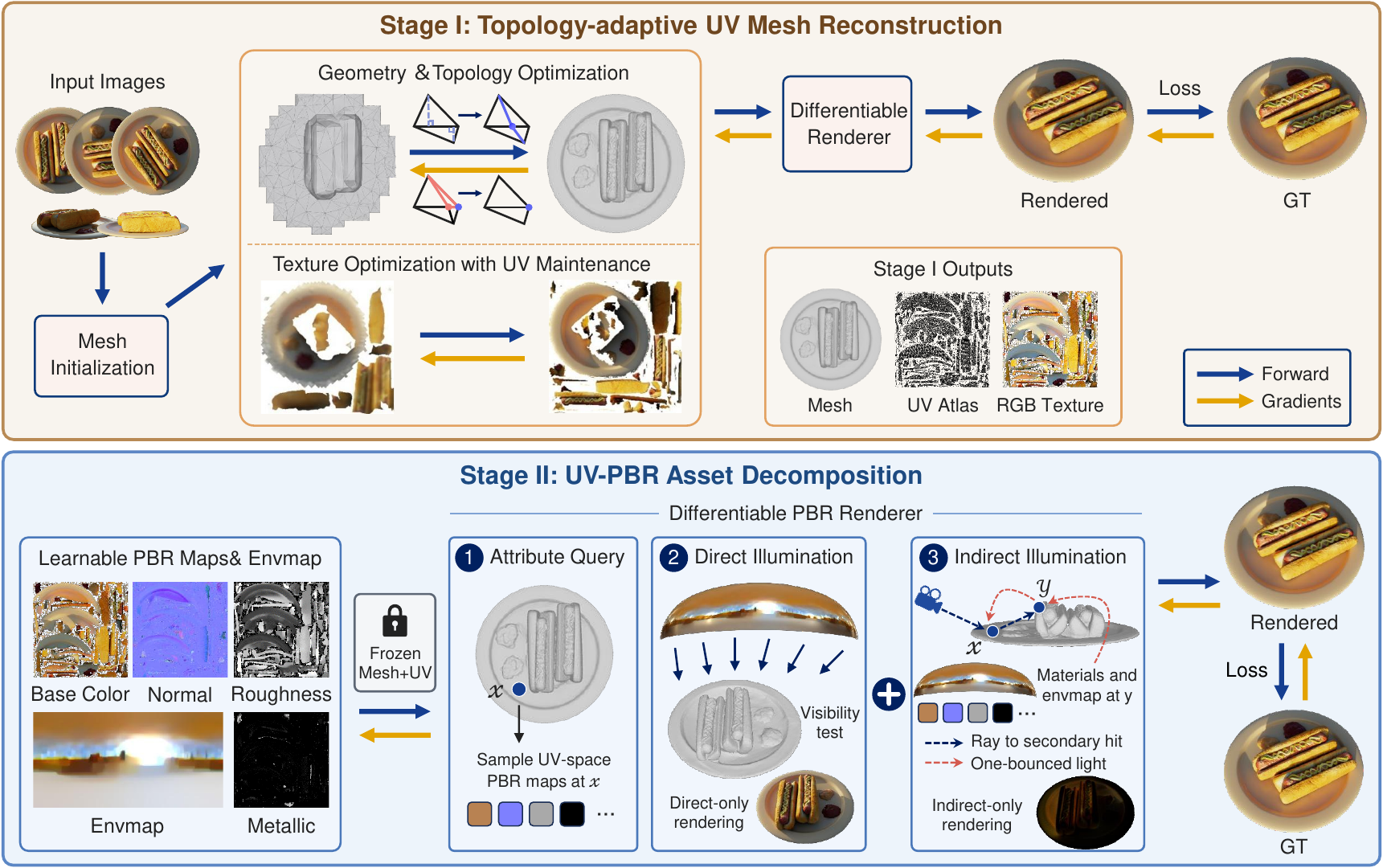}
  \vspace{-5mm}
  \caption{\added{Overview of ExMesh++. Stage I reconstructs a stable mesh-UV carrier from multi-view images by jointly optimizing geometry, topology, and an RGB texture, with UV updates during vertex splitting and merging. Stage II freezes the reconstructed mesh and UV coordinates, then optimizes UV-space PBR maps and environment lighting with a differentiable renderer. Material attributes are queried from the shared UV maps for both primary and secondary surface points, enabling direct illumination and one-bounce diffuse indirect illumination on the explicit mesh.}}
  \label{fig:method-overview}
\end{figure*}

\begin{addedblock}
\subsection{PBR Inverse Rendering}

\textbf{Neural inverse rendering.} Neural inverse rendering decomposes multi-view observations into geometry, materials, and illumination for relighting. NeRFactor~\cite{zhang2021nerfactor} uses neural fields to represent reflectance, visibility, and incident illumination. InvRender~\cite{zhang2022invrender} derives spatially varying indirect illumination from a learned radiance field. NeILF~\cite{yao2022neilf} and NeILF++~\cite{zhang2023neilfpp} model complex lighting with neural incident light fields, while TensoIR~\cite{jin2023tensoir} improves efficiency through tensor factorization. These methods can represent shadows, occlusion, and spatially varying illumination. However, their flexible representations leave considerable ambiguity among materials, visibility, and lighting. Appearance is also queried through neural or tensor fields, making the recovered components less convenient to edit than standard PBR maps.

\textbf{Gaussian-based inverse rendering.} Recent methods attach normals, albedo, roughness, visibility, and incident illumination to 3D Gaussian primitives to combine inverse-rendering quality with efficient rendering. GS-IR~\cite{liang2024gsir} uses depth-derived normals and baked volumes to approximate occlusion and indirect illumination. Relightable 3D Gaussians~\cite{gao2024relightable3dgs} combines BRDF decomposition with point-based ray tracing, while GeoSplatting~\cite{ye2025geosplatting} introduces mesh geometry to improve surface modeling and normal estimation. SVG-IR~\cite{SVGIR2025} further introduces spatially varying Gaussian attributes and a physically based indirect-lighting model. These methods inherit the efficient optimization and rendering of Gaussian representations. However, their material parameters remain attached to discrete Gaussians or Gaussian--mesh hybrids rather than a unified UV parameterization.

\textbf{Mesh-based PBR inverse rendering.} NVDiffRec~\cite{munkberg2022nvdiffrec} jointly optimizes topology, materials, and environment lighting with DMTet and exports a triangle mesh with texture maps. NVDiffRecMC~\cite{hasselgren2022nvdiffrecmc} extends this formulation with Monte Carlo direct lighting and multiple importance sampling, providing a more realistic treatment of shadows and specular reflection. MIRReS~\cite{dai2025inverse} adopts a two-stage framework that first extracts an explicit triangular mesh and then jointly refines geometry, materials, and lighting using multi-bounce path tracing and reservoir sampling. These methods demonstrate the effectiveness of physically based inverse rendering for explicit asset reconstruction. However, NVDiffRec and NVDiffRecMC optimize topology through a tetrahedral SDF representation, while MIRReS continues to refine the explicit geometry during material-light decomposition. Joint optimization makes effective use of appearance cues, but under limited supervision it also allows these variables to compensate for one another. This motivates a staged formulation that separates carrier reconstruction from material-light decomposition.

\subsection{Indirect Light Transport}

Existing approaches to indirect illumination can be grouped into three categories. The first uses learned light or radiance fields. InvRender~\cite{zhang2022invrender} derives spatially varying indirect illumination from a learned radiance field, while NeILF++~\cite{zhang2023neilfpp} couples incident and outgoing neural light fields through interreflection constraints. TensoIR~\cite{jin2023tensoir} represents secondary shading effects with tensor-factorized neural fields. The second group stores approximate occlusion or indirect illumination in a scene-dependent cache. GS-IR~\cite{liang2024gsir}, for example, bakes these quantities into volumetric grids. The third group explicitly simulates light transfer between surfaces. NeFII~\cite{wu2023nefii} traces secondary rays and caches indirect radiance in a neural field. IRGS~\cite{gu2025irgs} computes interreflection with differentiable 2D Gaussian ray tracing, while RadiosityGS~\cite{jiang2025radiositygs} extends radiosity to Gaussian surfels. Within this last category, our method adopts a lightweight one-bounce diffuse formulation on the mesh to capture local color bleeding and illumination in occluded regions.
\end{addedblock}

\end{revisedblock}
\begin{addedblock}
\section{Overview}
\label{sec:overview}

As shown in Fig.~\ref{fig:method-overview}, ExMesh++ follows a two-stage asset reconstruction pipeline. Stage I builds on our prior ExMesh framework \cite{Fan2026ExMesh} to reconstruct a compact explicit mesh together with a valid UV parameterization and an RGB appearance texture. It alternates differentiable optimization of mesh geometry and texture with discrete vertex splitting and merging, while updating the UV mapping to remain consistent with topology changes. Stage II freezes the resulting mesh-UV carrier and decomposes its appearance into UV-space PBR maps and environment illumination. The fixed carrier is further used to compute one-bounce diffuse indirect illumination through secondary-ray tracing and shared UV-PBR material queries. Sec.~\ref{sec:reconstruction} details the topology-adaptive mesh-UV reconstruction, and Sec.~\ref{sec:decomposition} describes PBR decomposition and the one-bounce indirect-lighting formulation.

\end{addedblock}
\begin{revisedblock}
\section{Topology-Adaptive Mesh-UV Reconstruction}
\label{sec:reconstruction}

Given multi-view images and their camera parameters, the first stage constructs a stable mesh-UV asset carrier. We represent the explicit mesh as \(\mathcal{M}=(V,F)\), where \(V\) denotes the 3D vertex positions and \(F\) denotes the set of triangular faces. To decouple appearance resolution from geometric density, we maintain a set of UV coordinates \(U\) and a face-corner-to-UV index map \(\Phi\). The Stage-I RGB appearance is stored in a fixed-resolution UV texture map \(T_{\mathrm{rgb}}\). In practice, we first train PGSR~\cite{chen2024pgsr} for 5K iterations and extract a coarse mesh through TSDF fusion~\cite{curless1996volumetric}. We then build an initial UV parameterization for the coarse mesh and initialize the RGB UV texture map. Our topology operations follow the classical vertex-split and edge-collapse paradigm of progressive meshes~\cite{hoppe1996progressive}, with task-specific criteria and UV updates designed for image-supervised reconstruction. The following sections describe how the mesh topology, UV mapping, and texture are optimized in Stage I.

\begin{figure}[t]
  \Description{Illustration of vertex splitting and vertex merging operations on triangular mesh faces.}
  \centering
  \includegraphics[width=\columnwidth]{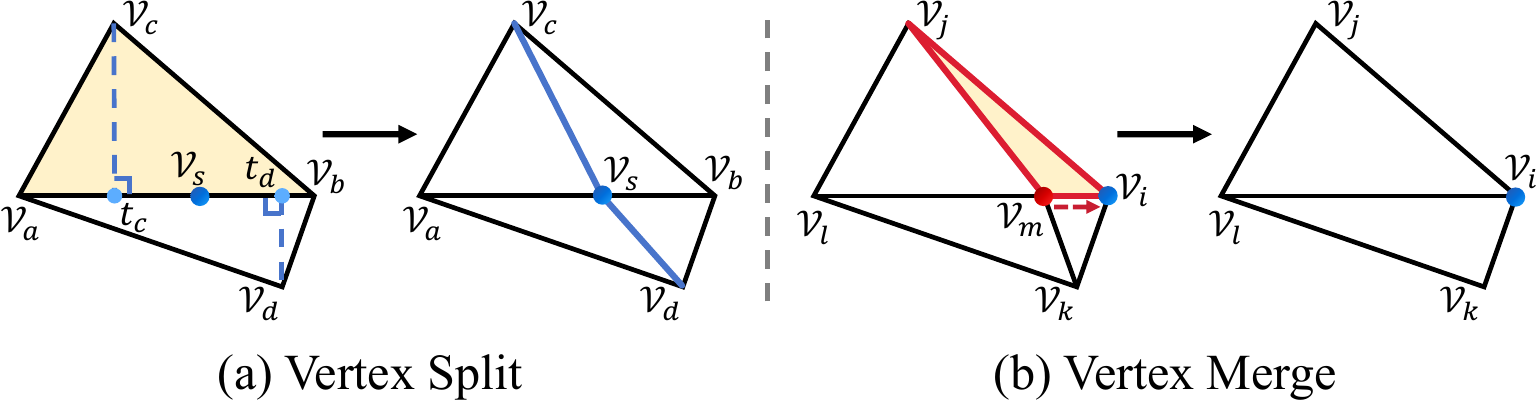}
  \caption{\revised{Discrete topological operations. (a) Vertex splitting inserts a new vertex on the selected edge and replaces the adjacent faces. (b) Vertex merging collapses an edge to simplify local topology.}}
  \label{fig:split-merge}
\end{figure}

\subsection{Vertex Splitting}
\label{sec:vertex-splitting}

\textbf{Splitting criterion.} Vertex splitting increases local geometric resolution in regions with complex geometry or large remaining optimization errors. We first maintain an exponential moving average of the position-gradient norm for each vertex. This EMA provides a temporal estimate of the optimization signal on each vertex:
\begin{equation}
\mathcal{G}_{v}^{(t)}
=
\beta_g \mathcal{G}_{v}^{(t-1)}
+
(1-\beta_g) \left\|\nabla_{v}\mathcal{L}^{(t)}\right\|_2 ,
\end{equation}
where \(t\) is the current iteration and \(\beta_g\) is the EMA coefficient. For each face \(f\), we define \(\mathcal{G}_f\) as the average gradient score of its three vertices. To capture local geometric variation, we estimate face curvature by averaging normal-angle differences with adjacent faces:
\begin{equation}
\mathcal{K}_{f}
=
\frac{1}{|\mathcal{N}(f)|}
\sum_{f' \in \mathcal{N}(f)}
\arccos
\left(
\mathbf{n}_{f} \cdot \mathbf{n}_{f'}
\right),
\end{equation}
where \(\mathcal{N}(f)\) is the set of faces adjacent to \(f\), and \(\mathbf{n}_{f}\) and \(\mathbf{n}_{f'}\) are their face normals. We restrict splitting candidates to relatively large faces, avoiding ineffective refinement in already dense regions. Each candidate face is assigned the following splitting score:
\begin{equation}
S_f = w_g \mathcal{G}_f + w_k \mathcal{K}_f ,
\end{equation}
where \(w_g\) and \(w_k\) balance the gradient-driven term and the curvature-driven term. Faces are selected for the current splitting round according to this score. This criterion combines optimization-driven density control with local geometric priors for adaptive explicit mesh refinement during reconstruction.

\textbf{Splitting operation.} For a selected triangular face \(f=(v_a,v_b,v_c)\), we first determine the edge to split. As shown in Fig.~\ref{fig:split-merge}(a), for each edge \(e=(v_i,v_j)\) of \(f\), let \(\ell_e\) be its length, and let \(d_e=(\deg(v_i)+\deg(v_j))/2\) be the average degree of its two endpoints. We define the edge score as \(S_e=\ell_e/d_e\), and split the edge with the highest score. This score favors long edges while penalizing highly connected endpoints, which helps reduce unstable topology changes around high-valence vertices.

Let the selected edge be \(e=(v_a,v_b)\). If \(e\) is an interior edge, it is shared by an adjacent face \(f'=(v_a,v_b,v_d)\). To better fit the local surface shape, we project \(v_c\) and \(v_d\) onto the line of \(e\), obtaining two projection points \(t_c\) and \(t_d\). The new vertex is initialized as \(v_s=(t_c+t_d)/2\). We then remove the two original faces \(f\) and \(f'\), and add four new faces: \((v_a,v_s,v_c)\), \((v_s,v_b,v_c)\), \((v_a,v_s,v_d)\), and \((v_s,v_b,v_d)\). If \(e\) is a boundary edge, there is no adjacent face \(f'\). In this case, we place \(v_s\) at the midpoint of \(e\), and replace the original face with two new faces. To avoid conflicts between concurrent topology changes, a splitting operation is skipped if any face to be modified has already been updated in the current round. The new vertex is further constrained to stay away from both edge endpoints. Specifically, \(\mu=\|v_s-v_a\|_2/\|v_b-v_a\|_2\) must satisfy \(\mu\in[0.25,0.75]\). Invalid splits are discarded to prevent skinny triangles.

\subsection{Vertex Merging}
\label{sec:vertex-merging}

\begin{figure}[t]
  \Description{Visualization showing vertex merging removing degenerate faces and producing cleaner local topology.}
  \centering
  \includegraphics[width=\columnwidth]{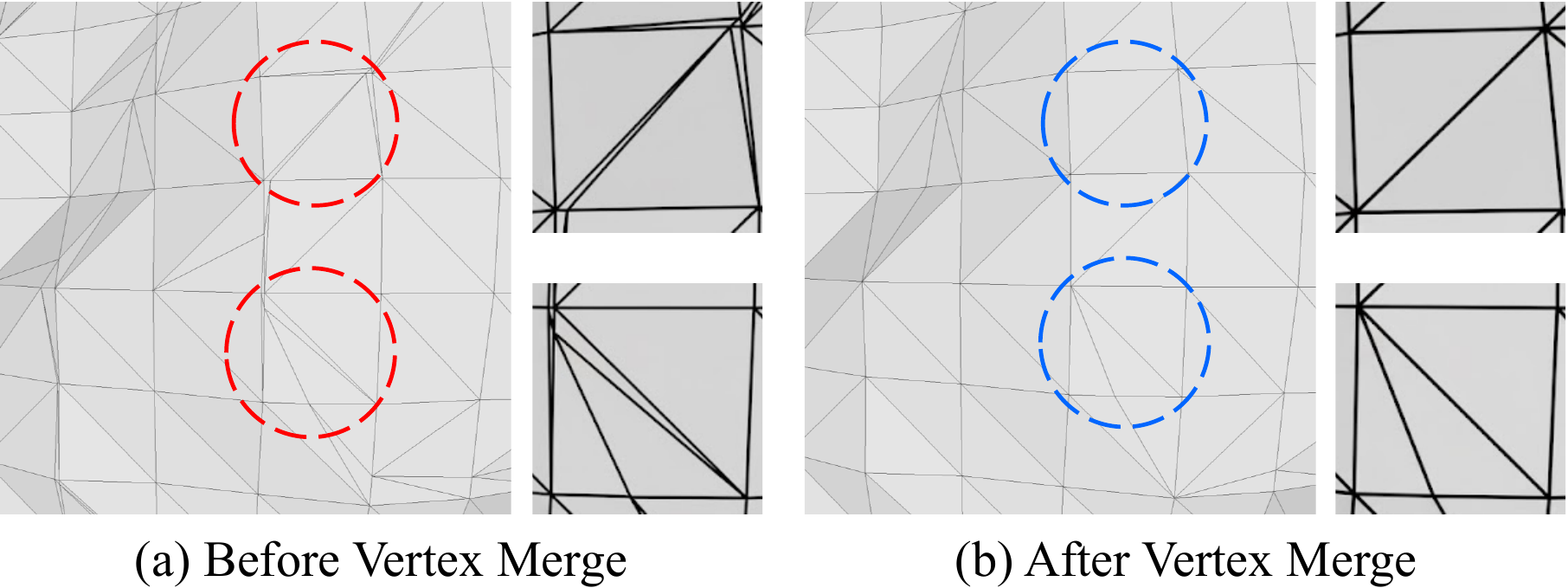}
  \caption{\textcolor{black}{Visualization of vertex merging. The merge operation eliminates degenerate faces and restores a cleaner local topology.}}
  \label{fig:vertex-merge-effect}
\end{figure}

\textbf{Merging criterion.} Vertex merging removes redundant structures and degenerate faces produced during optimization. It is implemented as a local edge-collapse operation~\cite{garland1997surface,hoppe1996progressive}. We select merging candidates using two cues: rendering visibility and geometric degeneracy. For rendering visibility, we record the contribution count \(C_{\mathrm{rend}}(f)\) of each face during rasterization over the training views. If a face is never rendered, i.e., \(C_{\mathrm{rend}}(f)=0\), it may be an internal face, a fully occluded face, or a redundant structure that does not contribute to the current reconstruction. To quantify geometric degeneracy, we use
\begin{equation}
\mathcal{D}_f
=
\frac{\mathrm{Area}(f)}
{\ell_{\max}^{2}(f)} ,
\end{equation}
where \(\mathrm{Area}(f)\) is the area of face \(f\), and \(\ell_{\max}(f)\) is the length of its longest edge. If \(\mathcal{D}_f < \tau_{\mathrm{degen}}\), the face is considered degenerate. To avoid over-simplifying large-scale structures, we only select merging candidates from faces with relatively small areas. Faces satisfying \(C_{\mathrm{rend}}(f)=0\) or \(\mathcal{D}_f < \tau_{\mathrm{degen}}\) are added to the merging set of the current round. Fig.~\ref{fig:vertex-merge-effect} illustrates this process, where vertex merging eliminates degenerate faces and restores a clean topology.

\textbf{Merging operation.} For a candidate face \(f=(v_a,v_b,v_c)\), we first determine which edge to collapse. As shown in Fig.~\ref{fig:split-merge}(b), if the face has a boundary edge, that edge is collapsed to preserve the object outline. If the face is an interior face, we collapse its shortest edge to preferentially remove fine degenerate structures. After choosing the collapse edge \(e=(v_i,v_j)\), we compare the degrees of its two endpoints. The endpoint with a lower degree is denoted as \(v_r\), and is merged into the other endpoint \(v_s\). The vertex \(v_r\) is then removed, and the adjacent faces are updated accordingly. Faces containing both \(v_r\) and \(v_s\) degenerate into edges after the collapse and are therefore deleted. Faces containing \(v_r\) but not \(v_s\) are reconnected by replacing \(v_r\) with \(v_s\). As in vertex splitting, the merge is skipped if any affected face has already been updated in the same round. This avoids conflicts between multiple discrete topology operations.

\subsection{UV Maintenance}
\label{sec:uv-maintenance}

Stage I changes the mesh connectivity during optimization, so the UV mapping must be updated together with each topology change. Each update keeps the face-corner UV index map \(\Phi\) valid after vertex insertion or deletion, while preserving separate UV coordinates along texture seams when needed. In this way, the RGB texture map \(T_{\mathrm{rgb}}\) remains a learnable appearance representation during topology adaptation, and the UV parameterization stays valid after vertex splitting or merging.

For vertex splitting, suppose the new vertex \(v_s\) lies on edge \(e=(v_a,v_b)\), with relative position \(\mu\). We interpolate the endpoint UV coordinates in the UV space of each related face. If \(e\) is a boundary edge, and the endpoint UVs in the original face are \(u_a\) and \(u_b\), then the UV coordinate of the new vertex is
\begin{equation}
u_s = (1-\mu)u_a + \mu u_b .
\end{equation}
This coordinate is assigned to all new face corners containing \(v_s\). If \(e\) is an interior edge, it belongs to two old faces. These two faces may lie on different UV islands. We therefore interpolate in the UV space of the two faces separately and obtain two candidate coordinates:
\begin{equation}
u_s^{(1)} = (1-\mu)u_a+\mu u_b,\quad
u_s^{(2)} = (1-\mu)u'_a+\mu u'_b .
\end{equation}
If \(\|u_s^{(1)}-u_s^{(2)}\|_2 < \tau_{\mathrm{uv}}\), the edge is treated as a non-seam edge. We average the two candidates, create one shared UV coordinate, and assign it to all related new faces. Otherwise, the edge is treated as crossing a UV seam. We create two independent UV coordinates for \(v_s\), and update the face-corner UV indices according to the original UV island of each new face. This seam-aware local update preserves the existing UV island boundaries during topology refinement.

For vertex merging, we update UV indices after deleting the vertex and reconnecting the adjacent faces. We first keep all UV coordinates that are still referenced by face corners, and remove UV coordinates that are no longer referenced by any face. We then build a mapping from old UV indices to new UV indices. This mapping is used to update all face-corner UV indices. As a result, the UV coordinate list remains compact, and every index in \(\Phi\) points to a valid UV coordinate. This operation only reorganizes indices, without moving UV coordinates still referenced by face corners.

\subsection{Stage-I Optimization}
\label{sec:stage-i-optimization}

Stage I uses differentiable rasterization to back-propagate image supervision to the explicit mesh and the UV texture map~\cite{laine2020modular}. At each iteration, we render the current mesh from a training view and obtain an RGB image \(\hat{I}\), an alpha mask \(\hat{S}\), and a depth map \(\hat{D}\). During rendering, the vertex positions \(V\) determine the mesh projection and visibility in the image plane. The face-corner UV coordinates are interpolated with barycentric coordinates to obtain the texture sampling location of each pixel, and the color is read from \(T_{\mathrm{rgb}}\). The Stage-I objective is defined as
\begin{equation}
\mathcal{L}_{\mathrm{I}}
=
\lambda_{\mathrm{rgb}}\mathcal{L}_{\mathrm{rgb}}
+
\lambda_d\mathcal{L}_d
+
\lambda_m\mathcal{L}_m
+
\lambda_s\mathcal{L}_s
+
\lambda_b\mathcal{L}_b .
\end{equation}
Here, \(\mathcal{L}_{\mathrm{rgb}}\) is the main image reconstruction loss. It combines an \(L_1\) term and a D-SSIM term~\cite{kerbl2023gaussian}:
\begin{equation}
\mathcal{L}_{\mathrm{rgb}}
=
(1-\lambda_{\mathrm{ssim}})
\|\hat{I}-I\|_1
+
\lambda_{\mathrm{ssim}}
\mathcal{L}_{\mathrm{D\text{-}SSIM}}(\hat{I},I).
\end{equation}
In addition to RGB supervision, we use a depth consistency loss \(\mathcal{L}_d\) and a silhouette loss \(\mathcal{L}_m\) to provide geometric cues. The depth consistency loss is computed inside the object mask. We obtain the reference depth \(D_{\mathrm{ref}}\) using Depth Anything 3~\cite{depthanything3} and measure its Pearson distance to the rendered depth \(\hat{D}\):
\begin{equation}
\mathcal{L}_{d}
=
1-
\frac{
\mathrm{Cov}(\hat{D},D_{\mathrm{ref}})
}{
\sigma_{\hat{D}}\sigma_{D_{\mathrm{ref}}}
}.
\end{equation}
The silhouette loss \(\mathcal{L}_m\) is the binary cross-entropy between the rendered alpha mask and the input mask. In addition, \(\mathcal{L}_s\) is a Laplacian smoothing regularizer that suppresses local high-frequency noise. \(\mathcal{L}_b\) is an adjacent-vertex offset regularizer that constrains local deformation consistency~\cite{munkberg2022nvdiffrec}.

After optimization, Stage I outputs a mesh \(\mathcal{M}^\ast\), a UV parameterization \((U^\ast,\Phi^\ast)\), and an RGB texture \(T_{\mathrm{rgb}}^\ast\). \(\mathcal{M}^\ast\), UV coordinates \(U^\ast\), and UV index map \(\Phi^\ast\) are fixed in Stage II. They serve as the asset carrier for UV-PBR material and lighting optimization.

\end{revisedblock}
\begin{addedblock}
\section{UV-PBR Asset Decomposition}
\label{sec:decomposition}

Given the mesh \(\mathcal{M}^\ast=(V^\ast,F^\ast)\), UV parameterization \((U^\ast,\Phi^\ast)\), and RGB texture \(T_{\mathrm{rgb}}^\ast\) from Stage I, the second stage fixes the vertex positions, mesh topology, and UV coordinates. It only optimizes PBR material maps and environment lighting in the shared UV space. 
Specifically, we represent the optimizable asset parameters as \(\Theta=\{A,R,M_{\mathrm{met}},N,E\}\). Here, \(A\) is a three-channel base-color map, \(R\) is a single-channel roughness map, \(M_{\mathrm{met}}\) is a single-channel metallic map, \(N\) is a tangent-space normal map, and \(E\) is a learnable RGB lat-long environment map. Note that the metallic map is optional. When metallic modeling is disabled, \(M_{\mathrm{met}}\) is omitted from optimization and \(m(x)\) is fixed to zero, yielding a dielectric material model. At the beginning of the PBR stage, \(A\) is initialized from the Stage-I RGB UV texture and converted from sRGB to linear RGB space. The roughness, normal, and environment maps are initialized to neutral values, while the optional metallic map is initialized to zero.

\subsection{Attribute Query}
\label{sec:attribute-query}

In PBR rendering, material parameters are queried from the UV space to 3D surfaces through the fixed mesh-UV carrier. For a visible intersection point \(x\) between a camera ray and the mesh, the renderer first obtains the face \(f\) containing \(x\) and its barycentric coordinates \(\mathbf{b}\). From \(\mathbf{b}\), we interpolate the surface position, the geometry normal \(\mathbf{n}_g(x)\), the view direction \(\omega_o\), and the UV coordinate \(u(x)\). We then use differentiable bilinear sampling to read material attributes from the PBR maps:
\begin{equation}
\begin{aligned}
a(x)&=A(u(x)),\quad
r(x)=R(u(x)),\\
m(x)&=M_{\mathrm{met}}(u(x)),\quad
q(x)=N(u(x)).
\end{aligned}
\end{equation}
Here, \(a(x)\), \(r(x)\), and \(m(x)\) denote base-color, roughness, and metallic, respectively. For the normal map, we first convert \(q(x)\) from texture values to a tangent-space normal. We then transform it to world space using the tangent frame interpolated on the current face, obtaining the shading normal \(\mathbf{n}_s(x)\). The PBR attributes at surface point \(x\) are denoted as
\begin{equation}
\mathcal{A}(x)=\{a(x),r(x),m(x),\mathbf{n}_s(x)\}.
\end{equation}
The same query applies to primary and secondary surface points.

\subsection{Direct Illumination}
\label{sec:direct-illumination}

For a primary surface point \(x\), we use the learnable environment map \(E\) to estimate its direct incident illumination. Given the view direction \(\omega_o\) and attributes \(\mathcal{A}(x)\), the direct outgoing radiance is
\begin{equation}
\begin{aligned}
L_{\mathrm{dir}}(x,\omega_o)
={}&
\int_{\Omega^+}
E(\omega_i)\,
\operatorname{Vis}(x,\omega_i)\,\\
&\quad
f_r(\mathcal{A}(x),\omega_i,\omega_o)\,
\max(\mathbf{n}_s(x)\cdot \omega_i,0)
\,d\omega_i ,
\end{aligned}
\end{equation}
where \(\Omega^+\) is the upper hemisphere around the shading normal, \(\omega_i\) is the incident direction, and \(\operatorname{Vis}(x,\omega_i)\) is the visibility from \(x\) along \(\omega_i\). We use a metallic-roughness PBR BRDF and decompose reflection into diffuse and specular components:
\begin{equation}
f_r(\mathcal{A}(x),\omega_i,\omega_o)
=
f_d(\mathcal{A}(x))
+
f_s(\mathcal{A}(x),\omega_i,\omega_o).
\end{equation}
Only the non-metallic component contributes to the diffuse term:
\begin{equation}
f_d(\mathcal{A}(x))
=
\frac{(1-m(x))\,a(x)}{\pi}.
\end{equation}
When \(m(x)=0\), the diffuse component is active, while the specular component uses the dielectric base reflectance. As \(m(x)\) approaches 1, the diffuse component vanishes, and the base-color determines the specular color. The specular term follows a microfacet BRDF~\cite{cook1982reflectance,walter2007microfacet,karis2013real}:
\begin{equation}
f_s
=
\frac{
D(\mathbf{h},\alpha)\,
G(\omega_i,\omega_o,\alpha)\,
F(\omega_i,\mathbf{h},F_0)
}{
4\,
\max(\mathbf{n}_s\cdot\omega_i,0)\,
\max(\mathbf{n}_s\cdot\omega_o,0)
+\epsilon
},
\end{equation}
where \(\mathbf{h}\) is the half vector, and \(\alpha\), derived from the roughness \(r(x)\), controls the width of the specular lobe. \(D\), \(G\), and \(F\) are the normal distribution function, geometry term, and Fresnel term, respectively. The Fresnel base reflectance \(F_0\) is interpolated between the dielectric reflectance \(F_{\mathrm{dielectric}}\) and the base-color according to metallic:
\begin{equation}
F_0=(1-m(x))F_{\mathrm{dielectric}}+m(x)a(x).
\end{equation}
Roughness controls the microfacet distribution and therefore the concentration of specular highlights.

During optimization, the above integral is approximated with Monte Carlo sampling. We build an importance sampling distribution from the brightness of the environment map and the spherical area weights, and sample incident directions from this distribution. For each sampled direction, ray tracing is used to test whether it is occluded by the reconstructed mesh~\cite{parker2010optix}. The learnable environment map is parameterized to be non-negative, so that radiance values remain valid. With \(N_d\) environment samples, the direct-light estimator is
\begin{equation}
\begin{aligned}
\hat{L}_{\mathrm{dir}}(x,\omega_o)
={}&
\frac{1}{N_d}
\sum_{k=1}^{N_d}
\frac{
E(\omega_i^k)\,
\operatorname{Vis}(x,\omega_i^k)\,
f_r(\mathcal{A}(x),\omega_i^k,\omega_o)
}{
p(\omega_i^k)
}\\
&\cdot
\max(\mathbf{n}_s(x)\cdot \omega_i^k,0)
,
\end{aligned}
\end{equation}
where \(p(\omega_i^k)\) is the sampling density. The rendered linear RGB result is used for log-\(L_1\) loss and is converted to sRGB for the D-SSIM term.

\subsection{One-Bounce Indirect Illumination}
\label{sec:one-bounce}

In addition to direct environment illumination, we explicitly compute one-bounce diffuse indirect illumination. For a primary surface point \(x\), the renderer samples a set of secondary directions \(\{\omega_j\}_{j=1}^{N_b}\) in the hemisphere defined by the geometry normal \(\mathbf{n}_g(x)\), and traces the corresponding rays on the fixed mesh. A secondary sample is considered valid only if the ray hits a front-facing surface, producing a secondary intersection point \(y_j\). For each valid \(y_j\), we obtain \(\mathcal{A}(y_j)\) using the same attribute query as in Sec.~\ref{sec:attribute-query}, based on its face index and barycentric coordinates.

At each secondary hit point, we estimate only the diffuse component under direct environment lighting, without modeling indirect specular reflection. Let \(f_d(y)=(1-m(y))a(y)/\pi\) be the diffuse BRDF term at the secondary point. The direct diffuse radiance at the secondary point is
\begin{equation}
L_{\mathrm{sec}}(y)
=
\int_{\Omega^+}
E(\omega_i)\,
\operatorname{Vis}(y,\omega_i)\,
f_d(y)\,
\max(\mathbf{n}_s(y)\cdot \omega_i,0)
\,d\omega_i .
\end{equation}
The returned radiance is then weighted by the diffuse BRDF term at the primary point \(x\). Using directions sampled over the primary hemisphere, we obtain the following Monte Carlo estimator for one-bounce indirect illumination:
\begin{equation}
\hat{L}_{\mathrm{ind}}(x)
=
\frac{1}{N_b}
\sum_{j=1}^{N_b}
\mathbf{1}_{\mathrm{hit}}(x,\omega_j)
\frac{
f_d(x)\,
L_{\mathrm{sec}}(y_j)\,
\max(\mathbf{n}_s(x)\cdot\omega_j,0)
}{
p_b(\omega_j)
},
\end{equation}
where \(p_b\) is the sampling probability density of secondary directions, and \(\mathbf{1}_{\mathrm{hit}}\) indicates whether the secondary ray produces a valid hit. The final rendered result for training is
\begin{equation}
\hat{I}=\hat{L}_{\mathrm{dir}}+\lambda_{\mathrm{ind}}\hat{L}_{\mathrm{ind}},
\end{equation}
where \(\lambda_{\mathrm{ind}}\) controls the strength of indirect illumination. Since the PBR materials and environment lighting are unstable at the beginning of Stage II, \(\lambda_{\mathrm{ind}}\) is gradually increased from 0 to its target value. This schedule allows the direct-light decomposition to reach an initial stable state before introducing color transfer between surfaces. The one-bounce term is computed entirely from the fixed mesh, the shared UV-PBR maps, and the environment map, without introducing an additional learned residual color field.

\subsection{Stage-II Optimization}
\label{sec:stage-ii-optimization}

For each training view, the fixed mesh and current PBR parameters are used to render an image \(\hat{I}\). We minimize the reconstruction error between \(\hat{I}\) and the ground-truth image \(I\). The image loss consists of a log-\(L_1\) term in linear RGB space and an SSIM term in sRGB space:
\begin{equation}
\mathcal{L}_{\mathrm{img}}
=
\lambda_{\log}
\left\|
\log(1+\hat{I})-\log(1+I)
\right\|_1
+
\lambda_{\mathrm{ssim}}
\mathcal{L}_{\mathrm{D\text{-}SSIM}}
(\Gamma(\hat{I}),\Gamma(I)),
\end{equation}
where \(\Gamma(\cdot)\) converts linear RGB to sRGB.

Using only the image reconstruction loss leaves ambiguities in material-light decomposition. We therefore introduce regularization on the projected material attributes and the environment map. The Stage-II objective is
\setcounter{equation}{21}
\begin{equation}
\mathcal{L}_{\mathrm{II}}
=
\mathcal{L}_{\mathrm{img}}
+\lambda_{\mathrm{mat}}\mathcal{L}_{\mathrm{mat}}
+\lambda_{\mathrm{env}}\mathcal{L}_{\mathrm{env}}
+\lambda_{\mathrm{chroma}}(t)\mathcal{L}_{\mathrm{chroma}}.
\end{equation}
For the current training view, let \(\widehat{P}\) denote the screen-space projection of a UV-space material map \(P\). We define the material and environment regularizers as
\begin{equation}
\mathcal{L}_{\mathrm{mat}}
=
\sum_{P\in\mathcal{P}_{mat}}
\operatorname{TV}(\widehat{P}),
\qquad
\mathcal{L}_{\mathrm{env}}
=
\operatorname{TV}(E),
\end{equation}
where \(\mathcal{P}_{mat}=\{A,R,N\}\) contains the base-color, roughness, and normal maps, and additionally includes \(M_{\mathrm{met}}\) when metallic optimization is enabled. Here, \(\operatorname{TV}(\cdot)\) denotes total variation over neighboring pixels. Accordingly, \(\mathcal{L}_{\mathrm{mat}}\) encourages locally smooth material attributes over adjacent visible surface points in image space, whereas \(\mathcal{L}_{\mathrm{env}}\) suppresses high-frequency noise in the environment map.

We additionally regularize the global color of the environment map during early optimization. Let \(\bar{\mathbf{E}}=(\bar{E}_r,\bar{E}_g,\bar{E}_b)\) denote its mean RGB radiance and \(\mu_E=(\bar{E}_r+\bar{E}_g+\bar{E}_b)/3\). The chroma regularizer is
\begin{equation}
\mathcal{L}_{\mathrm{chroma}}
=
\left\|
\bar{\mathbf{E}}-\mu_E\mathbf{1}
\right\|_2^2,
\qquad
\lambda_{\mathrm{chroma}}(t)
=
\lambda_{\mathrm{chroma}}^{0}
\left(1-\frac{t}{T}\right).
\end{equation}
This term discourages a global color cast in the recovered illumination. Its weight decreases linearly throughout training and reaches zero at the end of training.

\end{addedblock}
\section{Experiments}
\label{sec:experiments}

\begin{addedblock}

\begin{table}[t]
  \caption{\added{Default hyperparameter settings used in our implementation.}}
  \label{tab:hyperparameters}
  \centering
  \footnotesize
  \setlength{\tabcolsep}{1pt}
  \renewcommand{\arraystretch}{1.1}
  \resizebox{\columnwidth}{!}{%
  \begin{tabular}{@{}llr@{}}
    \toprule
     & Hyperparameter & Value \\
    \midrule
    \multirow{3}{*}{\begin{tabular}{l}Shared\\parameters\end{tabular}}
      & Texture/material map resolution & $2048\times2048$ \\
      & Texture/material learning rate & $2.5\times10^{-3}$ \\
      & D-SSIM weight $\lambda_{\mathrm{ssim}}$ & 0.2 \\
    \midrule
    \multirow{9}{*}{\begin{tabular}{l}Stage I\end{tabular}}
      & Initial vertex learning rate & $5\times10^{-4}$ \\
      & Gradient EMA coefficient $\beta_g$ & 0.95 \\
      & Split-score weights $w_g, w_k$ & 0.6, 0.4 \\
      & Maximum face count & 200K \\
      & Degeneracy threshold $\tau_{\mathrm{degen}}$ & 0.05 \\
      & Depth weight $\lambda_d$ & 0.01 \\
      & Silhouette weight $\lambda_m$ & 0.005 \\
      & Smoothness weight $\lambda_s$ & 1000 \\
      & Deviation weight $\lambda_b$ & 1000 \\
    \midrule
    \multirow{7}{*}{\begin{tabular}{l}Stage II\end{tabular}}
      & Environment-map resolution & $256\times512$ \\
      & Environment learning rate & $1.5\times10^{-2}$ \\
      & Direct-light samples $N_d$ & 8 \\
      & Secondary-ray samples $N_b$ & 16 \\
      & Light samples per secondary hit $N_s$ & 4 \\
      & Material/environment TV weights $\lambda_{\mathrm{mat}},\lambda_{\mathrm{env}}$ & 0.1 \\
      & Initial chroma weight $\lambda_{\mathrm{chroma}}^0$ & 0.1 \\
      & Target indirect-light weight $\lambda_{\mathrm{ind}}^{\mathrm{max}}$ & 1.0 \\
    \bottomrule
  \end{tabular}}
\end{table}

\newcommand{\first}[1]{\cellcolor{colorfirst}#1}
\newcommand{\second}[1]{\cellcolor{colorsecond}#1}
\newcommand{\third}[1]{\cellcolor{colorthird}#1}

\begin{table*}[t]
  \caption{\revised{Quantitative geometry comparison on DTU, reporting per-scan and average Chamfer Distance, total runtime, and the number of mesh vertices. Our runtime includes 3 minutes of initialization and 10 minutes of Stage-I optimization. {\color{colorfirst}\raisebox{-0.15mm}{\rule{3mm}{2mm}}} denotes the best, {\color{colorsecond}\raisebox{-0.15mm}{\rule{3mm}{2mm}}} second best, {\color{colorthird}\raisebox{-0.15mm}{\rule{3mm}{2mm}}} third best, respectively.}}
  \label{tab:dtu-geometry}
  \centering
  \scriptsize
  \setlength{\tabcolsep}{2.5pt}
  \setlength{\arrayrulewidth}{0.3pt}
  \resizebox{\textwidth}{!}{%
  \begin{tabular}{llcccccccccccccccccc}
    \toprule
    & Method & 24 & 37 & 40 & 55 & 63 & 65 & 69 & 83 & 97 & 105 & 106 & 110 & 114 & 118 & 122 & Avg. & Time & \#V \\
    \midrule
    \multirow{3}{*}{\begin{tabular}{c}NeRF-\\based\end{tabular}}
    & VolSDF~\cite{yariv2021volsdf} & 1.14 & 1.26 & 0.81 & 0.49 & 1.25 & 0.70 & 0.72 & 1.29 & 1.18 & 0.70 & 0.66 & 1.08 & 0.42 & 0.61 & 0.55 & 0.86 & $>$12h & 1M \\
    & NeuS~\cite{wang2021neus} & 0.83 & 0.98 & 0.56 & 0.37 & 1.13 & 0.59 & 0.60 & 1.45 & 0.95 & 0.78 & 0.52 & 1.43 & 0.36 & 0.45 & 0.45 & 0.77 & $>$12h & 488K \\
    & Neuralangelo~\cite{tancik2023neuralangelo} & 0.45 & 0.74 & 0.33 & \third{0.34} & 1.05 & \second{0.54} & 0.53 & 1.33 & 1.05 & 0.72 & \second{0.43} & 0.69 & 0.34 & \third{0.38} & 0.42 & 0.62 & $>$12h & 1M \\
    \midrule
    \multirow{5}{*}{\begin{tabular}{c}GS-\\based\end{tabular}}
    & SuGaR~\cite{su2023sugar} & 1.47 & 1.33 & 1.13 & 0.61 & 2.25 & 1.71 & 1.15 & 1.63 & 1.62 & 1.07 & 0.79 & 2.45 & 0.98 & 0.88 & 0.79 & 1.33 & 1h & 492K \\
    & 2DGS~\cite{huang2024twodgs} & 0.46 & 0.84 & \third{0.31} & 0.45 & 0.92 & 1.01 & 0.83 & 1.23 & 1.30 & 0.66 & 0.61 & 1.07 & 0.45 & 0.71 & 0.54 & 0.76 & 11m & 134K \\
    & GOF~\cite{yu2024gof} & 0.50 & 0.82 & 0.37 & 0.37 & 1.12 & 0.74 & 0.73 & 1.18 & 1.29 & 0.68 & 0.77 & 0.90 & 0.42 & 0.66 & 0.49 & 0.74 & 1h & 532K \\
    & PGSR~\cite{chen2024pgsr} & \second{0.36} & \second{0.57} & 0.38 & \first{0.33} & 0.78 & 0.58 & \second{0.50} & \third{1.08} & \second{0.63} & \third{0.59} & 0.46 & \second{0.54} & \first{0.30} & \third{0.38} & \second{0.34} & \second{0.52} & 30m & 540K \\
    & QGS~\cite{zhang2025qgs} & \third{0.38} & \third{0.62} & 0.37 & 0.38 & \second{0.75} & \third{0.55} & \third{0.51} & 1.12 & \third{0.68} & 0.61 & 0.46 & \third{0.58} & 0.35 & 0.41 & 0.40 & \third{0.54} & 48m & 129K \\
    \midrule
    \multirow{4}{*}{\begin{tabular}{c}Mesh-\\driven\end{tabular}}
    & NVDiffRec~\cite{munkberg2022nvdiffrec} & 3.04 & 3.02 & 2.10 & 0.78 & 2.18 & 1.60 & 1.46 & 1.67 & 2.85 & 1.26 & 1.10 & 3.26 & 1.13 & 1.31 & 1.19 & 1.86 & $>$1h & 69K \\
    & IMLS-Splat~\cite{yang2025imlssplatting} & 0.32 & 1.32 & 0.67 & 0.62 & 1.16 & 0.80 & 0.78 & 1.45 & 1.06 & 0.89 & 0.67 & 0.97 & 0.63 & 0.63 & 0.64 & 0.84 & 15m & 159K \\
    & GeoSVR~\cite{li2025geosvr} & \first{0.32} & \first{0.51} & \first{0.30} & \first{0.33} & \first{0.71} & \first{0.48} & \first{0.42} & \first{1.03} & \first{0.62} & \first{0.56} & \first{0.33} & \first{0.46} & \first{0.30} & \first{0.34} & \first{0.32} & \first{0.47} & 49m & 489K \\
    & Ours (StageI) & 0.44 & 0.70 & \first{0.30} & \third{0.34} & \third{0.76} & 0.63 & 0.52 & \second{1.06} & 1.05 & \second{0.58} & \third{0.45} & 0.88 & \third{0.32} & \second{0.36} & \third{0.38} & 0.58 & 13m & 102K \\
    \bottomrule
  \end{tabular}}
\end{table*}

\begin{figure*}[t]
  \Description{Qualitative geometry comparison on the DTU dataset.}
  \centering
  \includegraphics[width=1.0\textwidth]{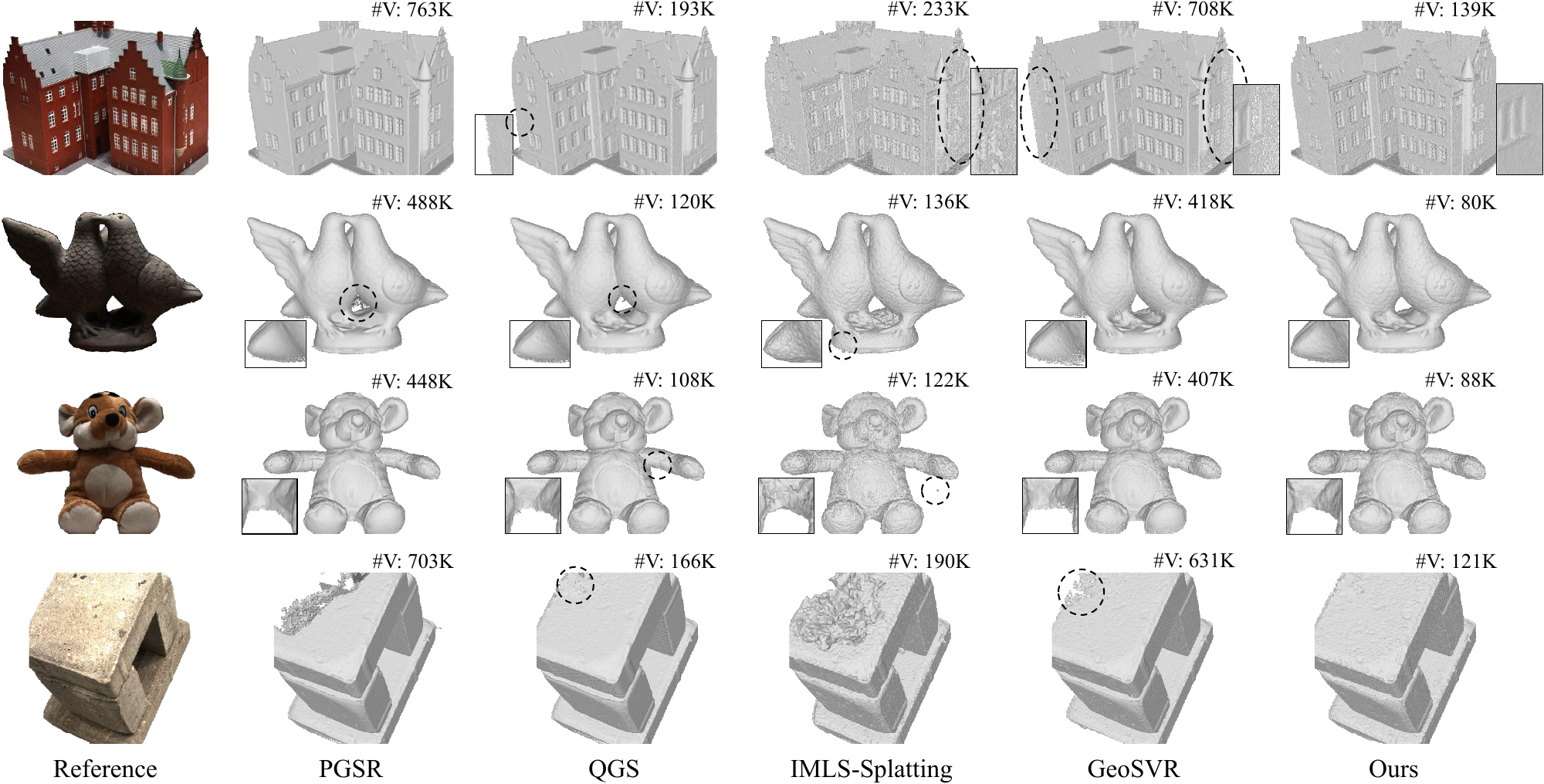}
  \caption{\revised{Qualitative geometry comparison on the DTU dataset, with number of vertices reported above each reconstruction. ExMesh++ recovers detailed and structurally complete surfaces while using fewer vertices and producing fewer floating fragments than the compared methods.}}
  \label{fig:dtu-geometry}
\end{figure*}

\begin{table*}[t]
  \caption{\added{Quantitative comparison on Synthetic4Relight in novel-view synthesis, relighting, material recovery, and runtime. Our runtime covers initialization and both training stages. ``Explicit Mesh'' indicates direct mesh output; ``Intermediate'' denotes meshes used internally but not exported as optimized assets.}}
  \label{tab:synthetic4relight}
  \centering
  \scriptsize
  \setlength{\tabcolsep}{2.5pt}
  \setlength{\arrayrulewidth}{0.3pt}
  \resizebox{\textwidth}{!}{%
  \begin{tabular}{l|c|ccc|ccc|ccc|c|c}
    \toprule
    \multirow{2}{*}{Method} & \multirow{2}{*}{Explicit Mesh} & \multicolumn{3}{c|}{Novel View Synthesis} & \multicolumn{3}{c|}{Relighting} & \multicolumn{3}{c|}{Albedo} & Roughness & \multirow{2}{*}{Time} \\[2pt]
    & & PSNR$\uparrow$ & SSIM$\uparrow$ & LPIPS$\downarrow$ & PSNR$\uparrow$ & SSIM$\uparrow$ & LPIPS$\downarrow$ & PSNR$\uparrow$ & SSIM$\uparrow$ & LPIPS$\downarrow$ & MSE$\downarrow$ & \\
    \midrule
    NeRFactor~\cite{zhang2021nerfactor} & \ding{55} & 22.80 & 0.916 & 0.150 & 21.54 & 0.875 & 0.171 & 19.49 & 0.864 & 0.206 & N/A & $>$48h \\
    NVDiffRecMC~\cite{hasselgren2022nvdiffrecmc} & \ding{51} & 34.29 & 0.967 & 0.068 & 24.22 & 0.943 & 0.078 & \third{29.61} & 0.945 & 0.075 & 0.009 & 2h \\
    InvRender~\cite{zhang2022invrender} & \ding{55} & 30.74 & 0.953 & 0.086 & 28.67 & 0.950 & 0.091 & 28.28 & 0.935 & 0.072 & \third{0.008} & $>$12h \\
    TensoIR~\cite{jin2023tensoir} & \ding{55} & 35.80 & 0.978 & \third{0.049} & 29.69 & 0.951 & 0.079 & \first{30.58} & 0.946 & 0.065 & 0.015 & $>$4h \\
    GS-IR~\cite{liang2024gsir} & \ding{55} & 35.65 & \third{0.971} & 0.055 & 27.93 & 0.953 & 0.067 & 20.78 & 0.907 & 0.101 & 0.045 & 19m \\
    RelightGS~\cite{gao2024relightable3dgs} & \ding{55} & \first{36.80} & \first{0.982} & \first{0.028} & \third{31.00} & \second{0.964} & \second{0.050} & 28.31 & \second{0.951} & 0.058 & 0.013 & 41m \\
    GeoSplatting~\cite{ye2025geosplatting} & intermediate & \second{35.99} & \second{0.978} & \second{0.031} & \second{34.10} & \first{0.971} & \first{0.037} & 29.90 & \third{0.949} & 0.062 & \first{0.004} & 27m \\
    Ours & \ding{51} & \third{35.76} & \third{0.971} & 0.053 & \first{34.19} & \third{0.962} & \third{0.061} & \second{30.16} & \first{0.953} & \second{0.068} & \second{0.005} & 30m \\
    \bottomrule
  \end{tabular}}
\end{table*}

\begin{figure*}[t]
  \Description{Synthetic4Relight visual comparison of material and lighting decomposition.}
  \centering
  \includegraphics[width=1.0\textwidth]{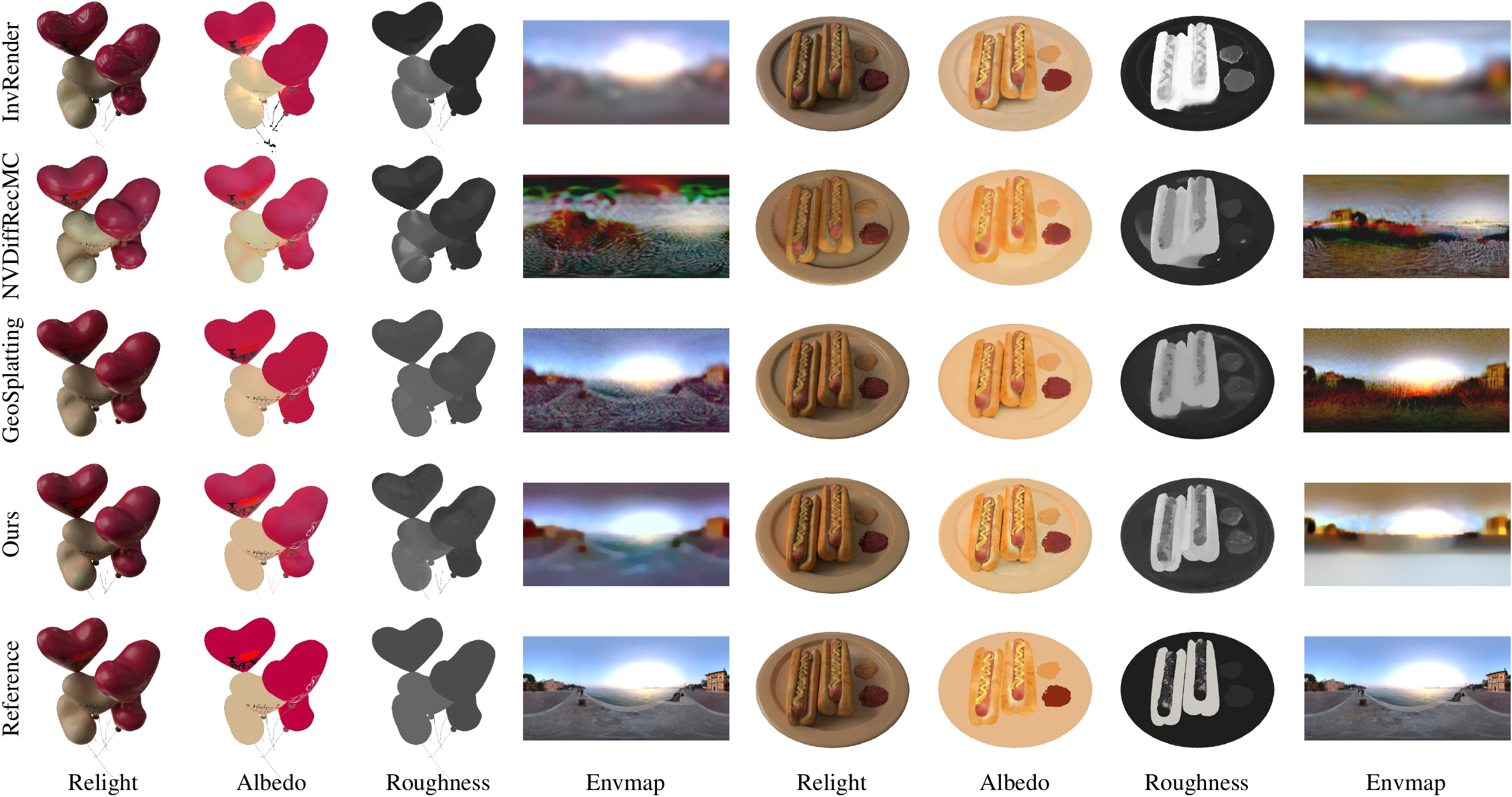}
  \caption{\added{Qualitative comparison of relighting and material-light decomposition on Synthetic4Relight. Each row shows the relit image, albedo, roughness, and recovered environment map. ExMesh++ better separates intrinsic materials from illumination while preserving plausible appearance under novel lighting.}}
  \label{fig:syn4rli-relight}
\end{figure*}

\begin{table*}[t]
  \caption{\added{Quantitative comparison on Stanford-ORB across novel-view synthesis, novel-scene relighting, geometry estimation, and runtime. PSNR-H and PSNR-L are evaluated in HDR linear space and tone-mapped LDR space, respectively.}}
  \label{tab:stanford-orb}
  \centering
  \scriptsize
  \setlength{\tabcolsep}{2.0pt}
  \setlength{\arrayrulewidth}{0.3pt}
  \resizebox{\textwidth}{!}{%
  \begin{tabular}{l|cccc|cccc|ccc|c}
    \toprule
    \multirow{2}{*}{Method} & \multicolumn{4}{c|}{Novel View Synthesis} & \multicolumn{4}{c|}{Novel Scene Relighting} & \multicolumn{3}{c|}{Geometry} & \multirow{2}{*}{Time} \\[2pt]
    & PSNR-H$\uparrow$ & PSNR-L$\uparrow$ & SSIM$\uparrow$ & LPIPS$\downarrow$ & PSNR-H$\uparrow$ & PSNR-L$\uparrow$ & SSIM$\uparrow$ & LPIPS$\downarrow$ & Depth$\downarrow$ & Normal$\downarrow$ & CD$\downarrow$ & \\
    \midrule
    Neuralangelo~\cite{tancik2023neuralangelo} & 27.69 & 35.87 & 0.981 & 0.029 & -- & -- & -- & -- & 69.68 & 1.85 & 77.85 & $>$12h \\
    2DGS~\cite{huang2024twodgs} & 29.63 & 37.30 & 0.984 & 0.022 & -- & -- & -- & -- & 28.43 & 0.11 & 1.35 & 11m \\
    PGSR~\cite{chen2024pgsr} & \third{30.77} & \first{40.52} & \second{0.989} & \first{0.012} & -- & -- & -- & -- & \third{0.36} & \third{0.05} & 0.73 & 30m \\
    \midrule
    NeRFactor~\cite{zhang2021nerfactor} & 26.06 & 33.47 & 0.973 & 0.046 & 23.54 & 30.38 & 0.969 & 0.048 & -- & -- & -- & $>$48h \\
    InvRender~\cite{zhang2022invrender} & 25.91 & 34.01 & 0.977 & 0.042 & 23.76 & 30.83 & 0.970 & 0.046 & -- & -- & -- & $>$12h \\
    GS-IR~\cite{liang2024gsir} & 26.48 & 32.66 & 0.960 & 0.050 & 22.88 & 29.05 & 0.958 & 0.053 & -- & -- & -- & 19m \\
    RelightGS~\cite{gao2024relightable3dgs} & 28.57 & 35.20 & 0.982 & 0.028 & 21.37 & 28.07 & 0.963 & 0.044 & -- & -- & -- & 41m \\
    NVDiffRecMC~\cite{hasselgren2022nvdiffrecmc} & 28.03 & 36.40 & 0.982 & 0.028 & \second{24.43} & \second{31.60} & \third{0.972} & 0.036 & \second{0.32} & \second{0.04} & 0.51 & 2h \\
    IRGS~\cite{gu2025irgs} & 28.82 & 35.66 & 0.978 & 0.034 & 21.94 & 28.68 & 0.969 & 0.039 & 0.54 & 0.06 & \third{0.49} & 40m \\
    RadiosityGS~\cite{jiang2025radiositygs} & \second{30.93} & \third{39.24} & \second{0.989} & \third{0.023} & \third{24.05} & \third{31.29} & \second{0.975} & \second{0.035} & 1.87 & 0.06 & \second{0.41} & 6h \\
    Ours & \first{31.27} & \second{39.52} & \first{0.990} & \second{0.014} & \first{26.60} & \first{34.03} & \first{0.980} & \first{0.022} & \first{0.30} & \first{0.02} & \first{0.30} & 30m \\
    \bottomrule
  \end{tabular}}
\end{table*}

\begin{figure*}[t]
  \Description{Stanford-ORB visual comparison on real-captured objects.}
  \centering
  \includegraphics[width=1.0\textwidth]{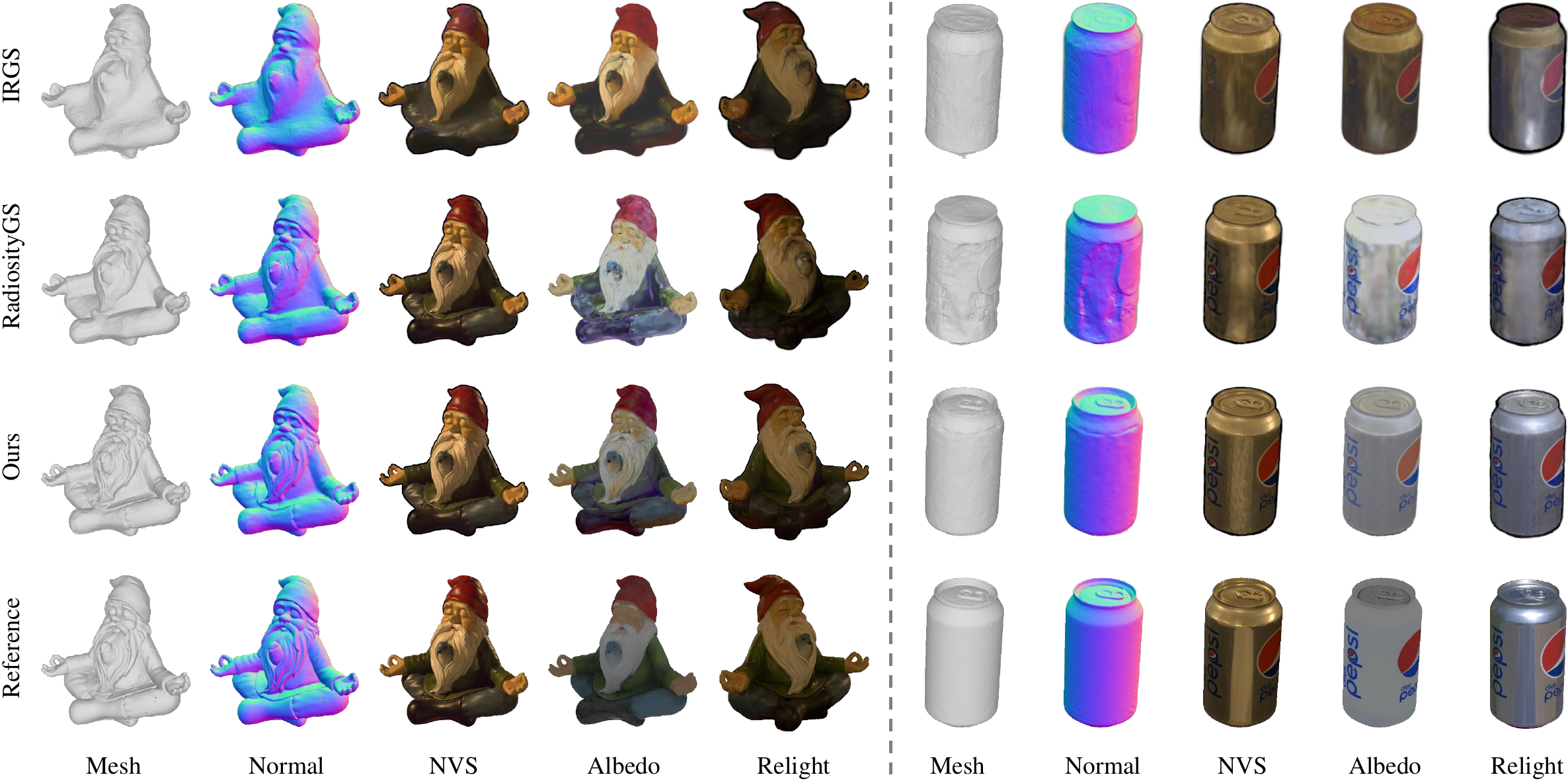}
  \caption{\added{Qualitative comparison on real-captured objects from Stanford-ORB. We compare reconstructed meshes, surface normals, novel-view renderings, albedo maps, and novel-scene relighting results. ExMesh++ achieves more consistent geometry, material decomposition, and relighting appearance.}}
  \label{fig:stanford-comprehensive}
\end{figure*}

\end{addedblock}

\subsection{Experimental Setup}
\label{sec:experimental-setup}

\begin{revisedblock}
\textbf{Implementation details.} 
We implement ExMesh++ in PyTorch and use nvdiffrast~\cite{laine2020modular} for differentiable rasterization. All experiments are conducted on a single NVIDIA RTX A6000 GPU. Unless otherwise stated, we use the same hyperparameter settings across all datasets and scenes, as summarized in Table~\ref{tab:hyperparameters}. For each scene, we first train PGSR~\cite{chen2024pgsr} for 5K iterations and extract an initial coarse mesh with TSDF fusion \cite{curless1996volumetric} at a resolution of $256^3$. We then build the initial UV parameterization and RGB UV texture. Stage I is optimized for 10K iterations. The first 1K iterations serve as a warm-up period without topology updates. From 1K to 6K iterations, vertex splitting and merging are performed every 500 iterations, with UV mapping updated accordingly. To prevent accumulated distortion and uneven texel allocation, we regenerate the UV atlas with xatlas every 2K iterations and transfer the current RGB texture to the new parameterization. After 6K iterations, the topology is fixed, while the vertex positions and RGB texture are refined until 10K iterations.

\end{revisedblock}

\begin{addedblock}
Stage II is trained for another 10K iterations. In this stage, vertex positions, topology, and UV coordinates are fixed, while base color, roughness, normal map, and a lat-long environment map are optimized. The PBR renderer approximates the environment-lighting integrals using Monte Carlo sampling~\cite{kajiya1986rendering} and evaluates visibility through OptiX ray tracing~\cite{parker2010optix}. At each primary surface point, we draw $N_d$ direct-light samples and trace $N_b$ secondary rays. Each valid secondary hit uses $N_s$ light samples to estimate its directly illuminated diffuse radiance. Secondary rays are evaluated at half resolution in both image dimensions, and the resulting indirect-light component is upsampled to full resolution. The indirect-light weight is linearly increased from 0 to 1 during the first 1K iterations of Stage II, allowing the direct material-light decomposition to stabilize before introducing interreflection.

Several baselines do not model metallic materials, and the evaluation datasets provide no reference metallic maps. For consistent benchmark comparisons, we therefore disable metallic optimization and fix the metallic map to zero. The optional metallic channel is retained for downstream asset editing.
\end{addedblock}

\begin{figure*}[t]
  \Description{DTU relighting comparison with one-bounce indirect lighting under training and novel environment maps.}
  \centering
  \includegraphics[width=0.95\textwidth]{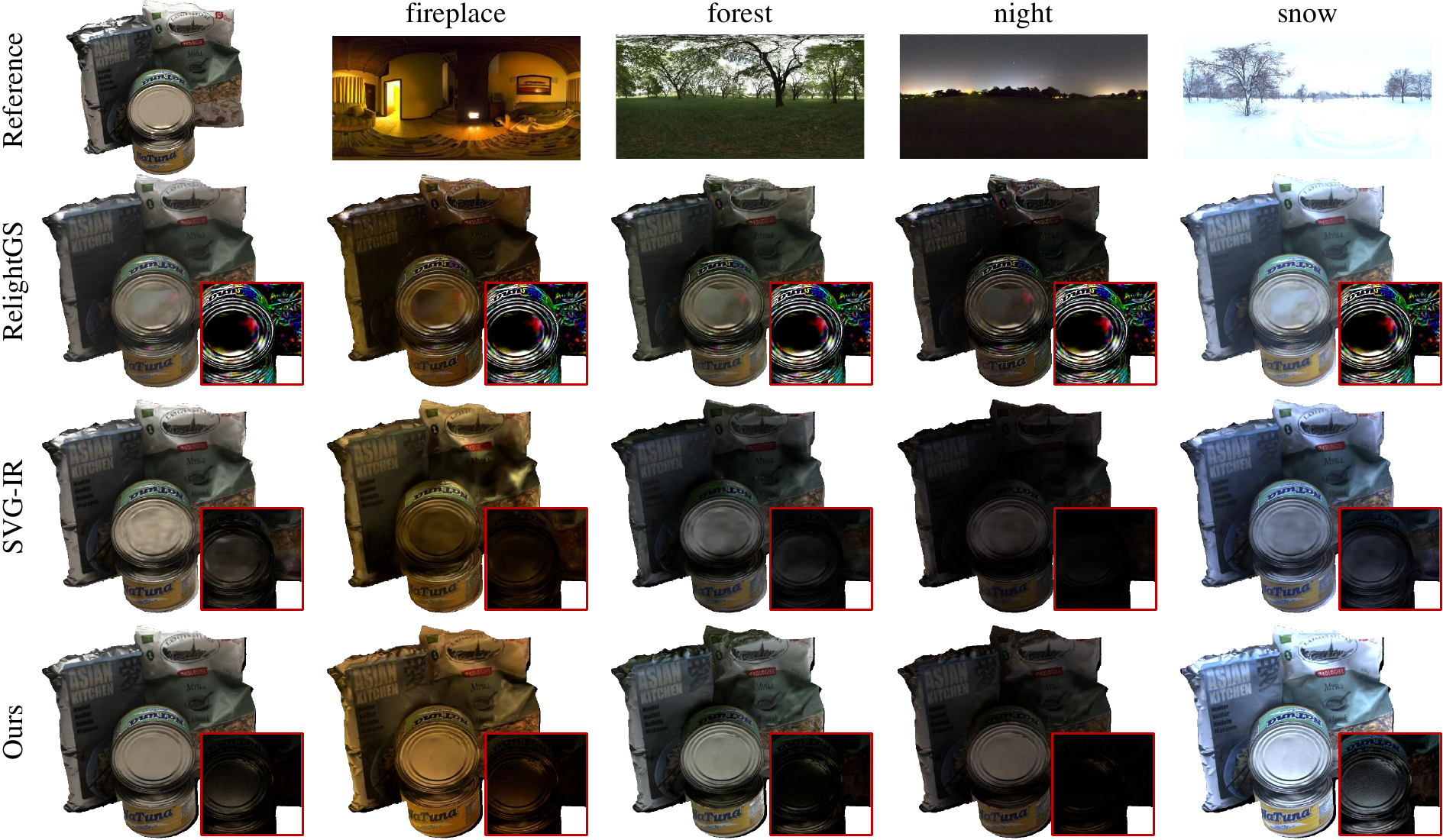}
  \caption{\added{Relighting comparison with indirect illumination on DTU scan97 under the training illumination and four novel environment maps. The red insets show the corresponding indirect-light components, highlighting the more continuous and environment-dependent light transfer produced by ExMesh++.}}
  \label{fig:dtu-relight}
\end{figure*}

\begin{addedblock}
\textbf{Datasets and Metrics.} We evaluate ExMesh++ on three datasets: DTU~\cite{jensen2014dtu}, Synthetic4Relight~\cite{zhang2022invrender}, and Stanford-ORB~\cite{kuang2023stanfordorb}. DTU contains 15 real-world objects, which are used for geometric evaluation and qualitative analysis of indirect lighting. Synthetic4Relight contains 4 CAD objects with self-occlusion and multiple materials, and provides test settings for novel-view synthesis, albedo, roughness, and novel-light relighting. Stanford-ORB contains captures of 14 real objects under 7 real environments, together with reference geometry, normal maps, and depth maps. We additionally use NeRF-Synthetic~\cite{mildenhall2020nerf} only for Stage-I ablations. Its controlled synthetic scenes and reference geometry allow these design choices to be evaluated without the ambiguity introduced by material-light decomposition.

On DTU, we report Chamfer Distance, runtime, and the number of mesh vertices. On Synthetic4Relight, we evaluate novel-view synthesis, relighting, and albedo with PSNR, SSIM~\cite{wang2004ssim}, and LPIPS~\cite{zhang2018lpips}, and evaluate roughness with MSE. Stanford-ORB provides both HDR and LDR versions of its captures. Following the benchmark protocol, all compared methods are trained on the HDR version. We report PSNR-H, PSNR-L, SSIM, LPIPS, depth error, normal error, and Chamfer Distance. PSNR-H is computed in linear HDR space, while PSNR-L, SSIM, and LPIPS are computed on the tone-mapped LDR images.

\textbf{Baselines.} For geometry reconstruction, we compare with three categories of surface reconstruction methods. NeRF-based methods include VolSDF~\cite{yariv2021volsdf}, NeuS~\cite{wang2021neus}, and Neuralangelo~\cite{tancik2023neuralangelo}. Gaussian-based methods include SuGaR~\cite{su2023sugar}, 2DGS~\cite{huang2024twodgs}, GOF~\cite{yu2024gof}, PGSR~\cite{chen2024pgsr}, and QGS~\cite{zhang2025qgs}. Mesh-driven methods include NVDiffRec~\cite{munkberg2022nvdiffrec}, IMLS-Splatting~\cite{yang2025imlssplatting}, and GeoSVR~\cite{li2025geosvr}. 

For relighting and material decomposition, we also compare with three categories of inverse rendering methods. Neural inverse rendering methods include NeRFactor~\cite{zhang2021nerfactor}, InvRender~\cite{zhang2022invrender}, and TensoIR~\cite{jin2023tensoir}. The explicit PBR baseline is NVDiffRecMC~\cite{hasselgren2022nvdiffrecmc}. Gaussian-based relighting methods include GS-IR~\cite{liang2024gsir}, RelightGS~\cite{gao2024relightable3dgs}, SVG-IR~\cite{SVGIR2025}, IRGS~\cite{gu2025irgs}, RadiosityGS~\cite{jiang2025radiositygs}, and GeoSplatting~\cite{ye2025geosplatting}.

\end{addedblock}

\subsection{Geometry Comparison}
\label{sec:geometry-comparison}

\begin{revisedblock}
We first evaluate geometry reconstruction quality on the DTU dataset. As shown in Table~\ref{tab:dtu-geometry}, ExMesh++ achieves an average Chamfer Distance of 0.58. It obtains geometry accuracy close to recent reconstruction methods, while requiring only 13 minutes of training time and a compact mesh representation. PGSR and GeoSVR achieve lower Chamfer Distance, but require longer training time and larger geometric representations. Mesh-driven methods such as NVDiffRec and IMLS-Splatting still show clear gaps in geometry accuracy or structural completeness. Fig.~\ref{fig:dtu-geometry} further illustrates the structural differences among different methods. PGSR and GeoSVR often produce redundant faces and local floating artifacts. QGS and IMLS-Splatting also show rough boundaries and isolated fragments. In contrast, ExMesh++ produces meshes without obvious artifacts, recovering details while maintaining smooth and intact surfaces.
\end{revisedblock}

\begin{figure}[t]
  \begin{addedblock}
  \Description{Blender relighting examples using exported ExMesh++ assets under different lighting conditions.}
  \centering
  \includegraphics[width=\columnwidth]{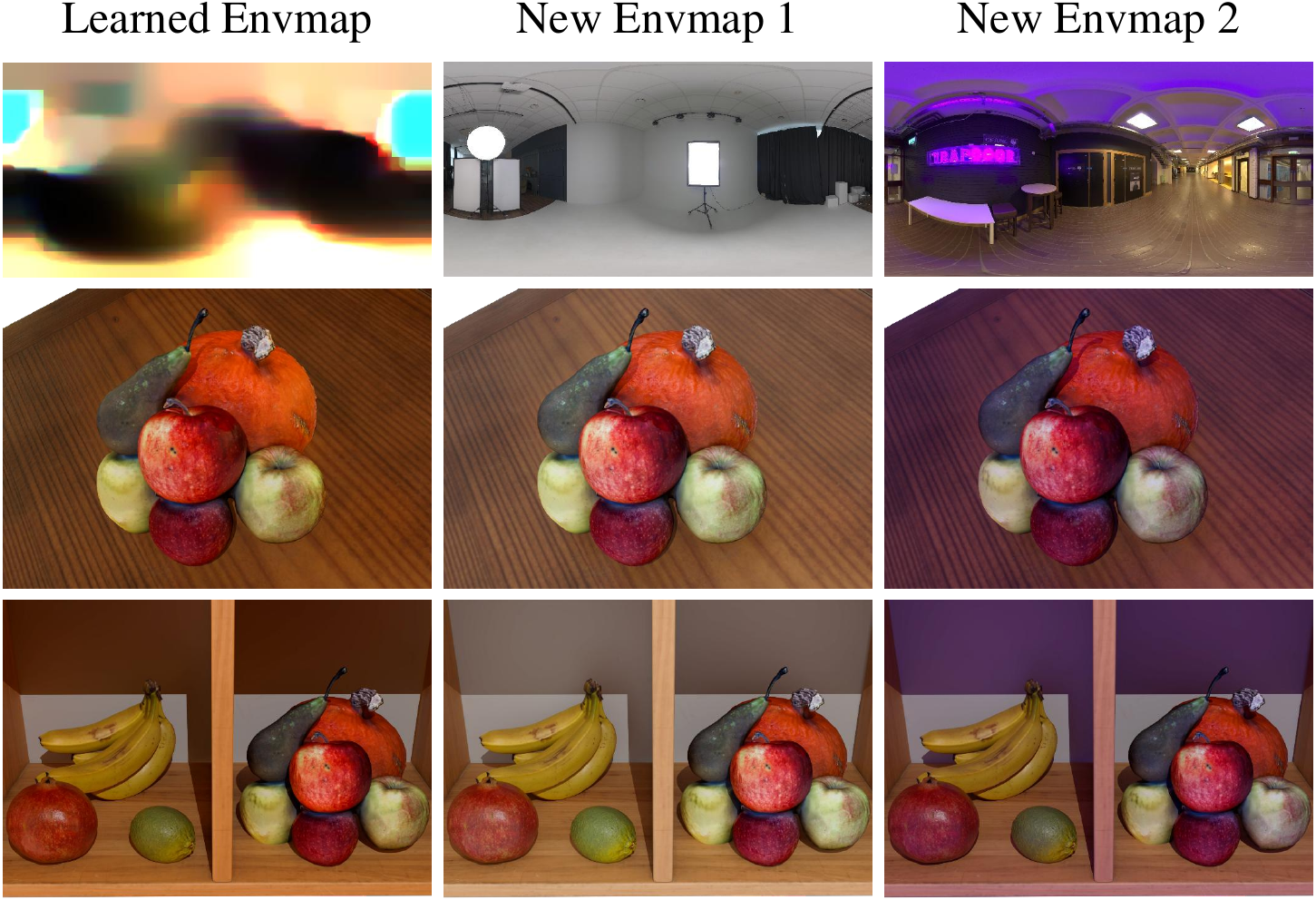}
  \caption{\added{Relighting of the exported UV-PBR asset in Blender under the recovered environment map and two novel environment maps. The middle row places the reconstructed object on a tabletop, while the bottom row combines it with artist-created assets.}}
  \label{fig:blender-relight}
  \end{addedblock}
\end{figure}

\subsection{Relighting Comparison}
\label{sec:relighting-comparison}

\begin{addedblock}
\textbf{Synthetic Evaluation.} We evaluate novel-view synthesis, relighting, and material decomposition on Synthetic4Relight. Table~\ref{tab:synthetic4relight} shows that our method achieves the highest relighting PSNR and competitive results on novel-view synthesis, albedo, and roughness estimation, while producing explicit mesh-UV-PBR assets. Fig.~\ref{fig:syn4rli-relight} further shows the differences in material and lighting decomposition. InvRender leaves visible shadow and highlight residuals in albedo, so its relighting results still contain traces of the original illumination. NVDiffRecMC produces noisy roughness maps, while its recovered environment maps show inaccurate light-source locations and global color tones. GeoSplatting produces noticeable high-frequency noise in the environment map. 

\textbf{Real-captured Evaluation.} We further evaluate novel-view synthesis, novel-scene relighting, and geometry on Stanford-ORB. Table~\ref{tab:stanford-orb} shows that ExMesh++ outperforms the compared methods on all novel-scene relighting metrics and achieves the lowest errors in depth, normal, and Chamfer Distance. Since PGSR, 2DGS, and Neuralangelo do not provide novel-scene relighting results, they are used as references for view synthesis and geometry reconstruction. Fig.~\ref{fig:stanford-comprehensive} shows visual comparisons on real-captured objects. IRGS produces over-smoothed meshes and leaves strong illumination colors in albedo, leading to a visible tone shift in relighting. RadiosityGS shows high-frequency noise in the mesh, normal, and albedo results. In comparison, ExMesh++ better matches the reference in geometry, albedo color distribution, and relighting appearance.
\end{addedblock}

\subsection{Indirect Lighting Comparison}
\label{sec:indirect-lighting-comparison}

\begin{addedblock}
We further compare relighting results with indirect illumination on DTU scan97. Fig.~\ref{fig:dtu-relight} shows relighting results under the training illumination and four novel environment maps. The red insets show the corresponding indirect-light components. RelightGS produces colorful high-frequency artifacts in the indirect-light component, which remain visible across different environment maps. SVG-IR captures the overall brightness change, but its indirect-light component is mostly concentrated in dark regions and shows weak local color transfer. ExMesh++ instead maintains environment-dependent illumination changes, with more continuous indirect-light variation around the can lid and occluded regions.

\begin{figure}[t]
  \Description{Texture and material editing examples of exported UV-PBR assets in Blender.}
  \centering
  \includegraphics[width=\columnwidth]{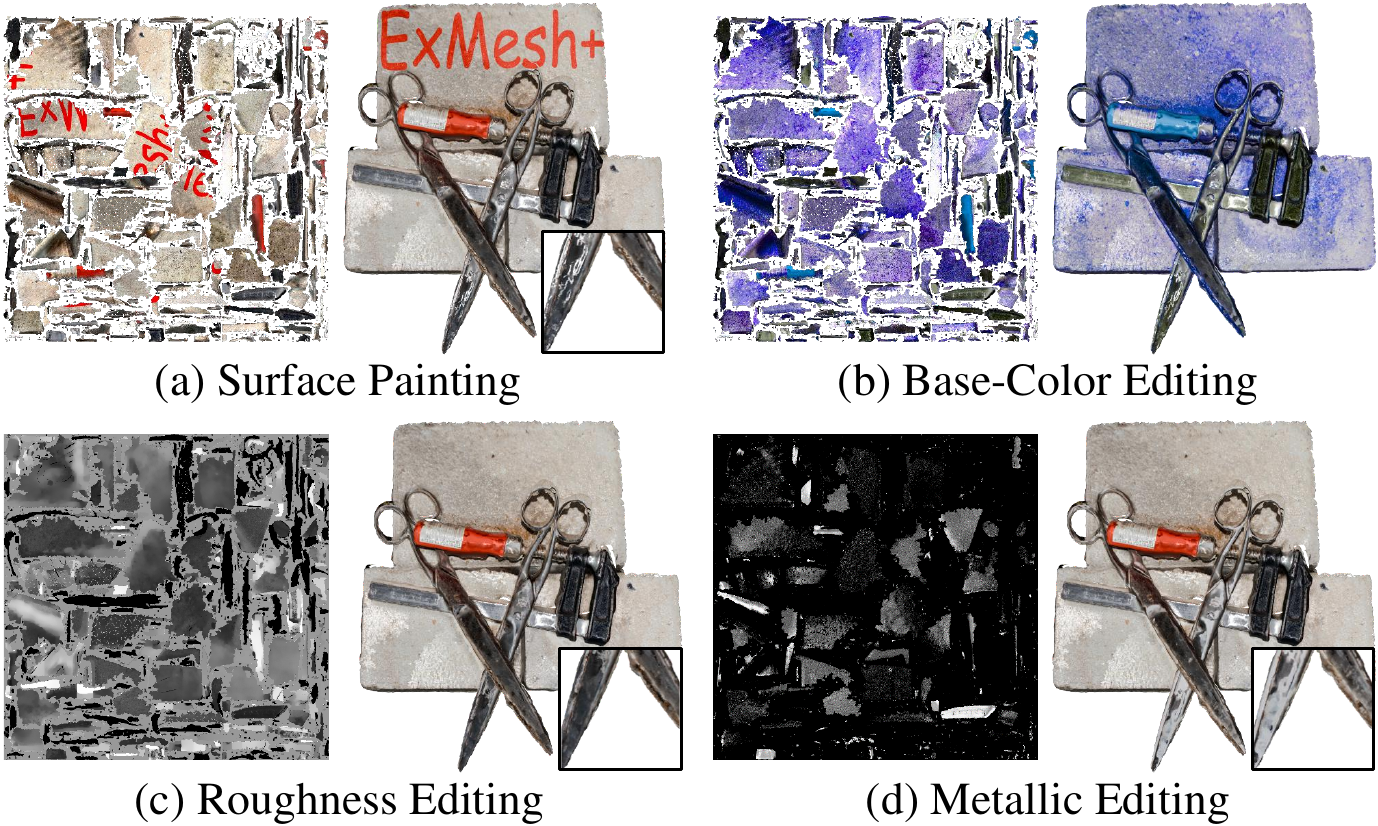}
  \caption{\added{Editing results of exported UV-PBR assets in Blender. The examples demonstrate direct surface painting, global base-color adjustment, and local edits to the roughness and metallic maps.}}
  \label{fig:material-edit}
\end{figure}

\subsection{Downstream Applications}
\label{sec:downstream-applications}

\textbf{DCC Relighting.} We import the mesh, UV-PBR materials, and environment lighting reconstructed from DTU scan63 into Blender to test the exported asset in a standard DCC workflow. Fig.~\ref{fig:blender-relight} shows the trained environment map and two new environment maps, together with Blender renderings under these lighting conditions. A weak point light is also added to produce visible shadows and local contact effects. The middle row shows the reconstructed object on a tabletop under the three lighting conditions. The bottom row places it together with artist-created fruit assets under the same lighting settings. Color changes and shadows of the reconstructed asset follow patterns similar to those of the artist-created assets.

\textbf{Texture Editing.} Fig.~\ref{fig:material-edit} shows texture and material editing results of exported UV-PBR assets in Blender. All results are rendered with a point light. Since ExMesh++ exports explicit UV textures and PBR material maps, different material channels can be directly edited in texture space. Fig.~\ref{fig:material-edit}(a) paints text on the object surface. Fig.~\ref{fig:material-edit}(b) edits the global color tone of the base-color map, changing the object surface from the original gray tone to a blue-purple tone while preserving geometry and local shading. Fig.~\ref{fig:material-edit}(c) and Fig.~\ref{fig:material-edit}(d) locally edit the roughness and metallic maps of the scissors region. After setting the roughness of the scissors region to 1, the scissors show a more matte appearance. After setting the metallic value of the scissors region to 0, the metallic highlights are clearly weakened, while unedited regions remain visually unchanged.
\end{addedblock}

\subsection{Computational Efficiency}
\label{sec:computational-efficiency}
\begin{addedblock}
We compare training resources, relighting speed, and output-mesh complexity on Stanford-ORB in Table~\ref{tab:efficiency}. Relighting speed is measured at the original resolution of $1600 \times 1600$, after model and camera initialization. The timing includes environment-map transfer and update, light-transport evaluation, shading, and image rendering, while excluding model loading, disk I/O, and image saving. We use the evaluation-quality configuration of each method. The first rendered frame is treated as warm-up and excluded, and FPS is averaged over the remaining frames. When a fixed environment map is shared by multiple views, its update cost is amortized across those views. Mesh sizes are measured using a unified PLY format.

The full ExMesh++ pipeline takes 30 minutes per object, including approximately 3 minutes for initialization, 10 minutes for Stage-I reconstruction, and 17 minutes for Stage-II optimization. Disabling indirect illumination reduces the Stage-II cost to 10 minutes, resulting in a total training time of 23 minutes. The direct-only variant renders at 40.47 FPS, while the full model with one-bounce indirect illumination achieves 15.09 FPS. Compared with prior mesh-producing inverse-rendering methods, ExMesh++ produces a compact output mesh while maintaining practical relighting speed.

\begin{table}[t]
  \caption{\added{Computational efficiency on Stanford-ORB.}}
  \label{tab:efficiency}
  \centering
  \scriptsize
  \setlength{\tabcolsep}{1.0pt}
  \renewcommand{\arraystretch}{1.1}
  \resizebox{\columnwidth}{!}{%
  \begin{tabular}{@{}lccccc@{}}
    \toprule
    Method & \begin{tabular}{c}Peak GPU\\Mem.\end{tabular} & \begin{tabular}{c}Training\\Time\end{tabular} & \begin{tabular}{c}Relighting\\FPS\end{tabular} & \#Vertices & \begin{tabular}{c}Mesh\\Size\end{tabular} \\
    \midrule
    NeRFactor & 16GB & $>$48h & 0.01 & -- & -- \\
    InvRender & 15GB & $>$12h & 0.03 & -- & -- \\
    GS-IR & 5GB & 19m & 54.50 & -- & -- \\
    RelightGS & 10GB & 41m & 10.49 & -- & -- \\
    NVDiffRecMC & 26GB & 2h & 4.55 & 47K & 2.48MB \\
    IRGS & 5GB & 40m & 0.19 & 289K & 14.56MB \\
    RadiosityGS & 34GB & 6h & 1.54 & 267K & 13.39MB \\
    \midrule
    Ours (direct only) & 10GB & 23m & 40.47 & 52K & 2.63MB \\
    Ours (full) & 10GB & 30m & 15.09 & 52K & 2.63MB \\
    \bottomrule
  \end{tabular}}
\end{table}
\end{addedblock}

\begin{table*}[t]
  \caption{\added{Quantitative ablations of topology operations, texture representations, PBR decomposition, and indirect illumination.}}
  \label{tab:ablation}
  \centering
  \begin{subtable}[t]{0.36\textwidth}
    \centering
    \normalsize
    \setlength{\tabcolsep}{2.5pt}
    \caption{\added{Stage-I ablations on NeRF-Synthetic.}}
    \begin{tabular*}{\linewidth}{@{\extracolsep{\fill}}lccc@{}}
      \toprule
      Setting & CD$\downarrow$ & PSNR$\uparrow$ & \#V \\
      \midrule
      Only Split & 0.74 & 28.77 & 121K \\
      Only Merge & 1.81 & 23.51 & 6K \\
      Random Split & 0.77 & 28.30 & 94K \\
      Random Merge & 1.34 & 26.27 & 98K \\
      \midrule
      Per-Vertex Color & 0.76 & 28.47 & 184K \\
      Per-Face Color & 0.78 & 28.12 & 227K \\
      MLP Texture & 0.64 & 28.87 & 101K \\
      \midrule
      Ours (full) & 0.64 & 29.32 & 100K \\
      \bottomrule
    \end{tabular*}
    \label{tab:ablation-topology-texture}
  \end{subtable}\hfill
  \begin{subtable}[t]{0.61\textwidth}
    \centering
    \normalsize
    \setlength{\tabcolsep}{2.5pt}
    \caption{\added{Stage-II ablations on Synthetic4Relight.}}
    \begin{tabular*}{\linewidth}{@{\extracolsep{\fill}}lcccccc@{}}
      \toprule
      \multirow{2}{*}{Setting} & \multicolumn{3}{c}{Relighting} & \multicolumn{3}{c}{Albedo} \\
      & PSNR$\uparrow$ & SSIM$\uparrow$ & LPIPS$\downarrow$ & PSNR$\uparrow$ & SSIM$\uparrow$ & LPIPS$\downarrow$ \\
      \midrule
      Stage-I Only & 27.74 & 0.918 & 0.065 & -- & -- & -- \\
      Gray Base-Color Init & 34.06 & 0.961 & 0.062 & 30.09 & 0.948 & 0.069 \\
      Stage-II Only & 29.55 & 0.927 & 0.062 & 27.20 & 0.921 & 0.092 \\
      \midrule
      Direct Only & 33.72 & 0.961 & 0.062 & 29.96 & 0.948 & 0.070 \\
      Eval-only Indirect & 34.00 & 0.962 & 0.062 & 29.99 & 0.948 & 0.069 \\
      No Secondary Material & 34.15 & 0.962 & 0.061 & 29.97 & 0.947 & 0.069 \\
      \midrule
      Ours (full) & 34.19 & 0.962 & 0.061 & 30.16 & 0.949 & 0.068 \\
      \bottomrule
    \end{tabular*}
    \label{tab:ablation-pbr-indirect}
  \end{subtable}
\end{table*}

\subsection{Ablation Study}
\label{sec:ablation-study}

\begin{revisedblock}
\textbf{Vertex Splitting and Merging.} We first analyze the effect of vertex splitting and merging on the NeRF-Synthetic dataset. Table~\ref{tab:ablation}(a) compares five variants: only split, only merge, random split, random merge, and the full method. In random split, the split edge is still selected by our criterion, but the new vertex is placed at a random position. In random merge, edges are collapsed randomly instead of using our merge criterion. Using only merging substantially reduces the number of vertices, but it cannot increase geometric resolution in complex regions. Using only splitting can add local details, but it also leads to more vertices. The two random variants underperform the full strategy, showing that both candidate selection and operation geometry affect local triangulation quality. Fig.~\ref{fig:ablation-split-merge} further shows that random splitting disrupts local mesh regularity, while random merging introduces unstable collapses. The full method preserves a compact mesh while producing a more continuous surface.

\textbf{Initialization Strategy.} Fig.~\ref{fig:ablation-wo-init} shows the random initialization experiment. We use a 1K-face sphere with random colors as the initial input, instead of a coarse mesh obtained from multi-view images. During optimization, the mesh gradually deforms from the sphere and increases its face count. It eventually forms the main structure of the target object. The UV texture also changes from random colors to a layout related to the object appearance. This experiment shows that the optimization process does not strictly depend on a specific initial mesh shape. However, compared with the standard initialization, starting from an unrelated sphere leads to slightly lower final geometry accuracy.

\textbf{Texture Representation.} We further compare UV texture with per-vertex color, per-face color, and coordinate-based MLP texture on NeRF-Synthetic. As shown in Table~\ref{tab:ablation}(a), per-vertex and per-face colors require many more vertices to represent high-frequency appearance, but still yield worse geometry and rendering quality. Coordinate-based MLP texture achieves comparable geometry accuracy, but does not provide a standard editable texture map. In comparison, UV texture decouples appearance resolution from mesh density and provides a shared texture space for PBR decomposition.

\begin{figure}[t]
  \Description{Ablation results for vertex splitting and merging strategies.}
  \centering
  \includegraphics[width=\columnwidth]{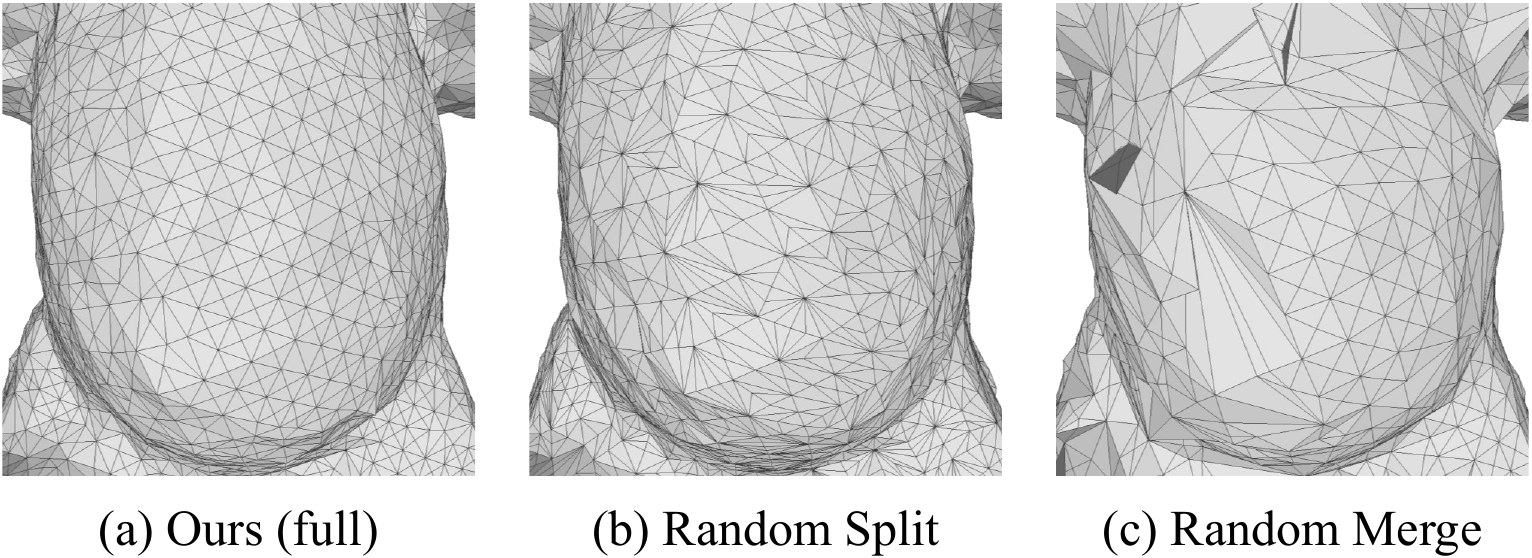}
  \caption{\textcolor{black}{Qualitative ablation of vertex splitting and merging strategies.}}
  \label{fig:ablation-split-merge}
\end{figure}

\begin{figure}[t]
  \Description{Random initialization experiment starting from a sphere with random colors.}
  \centering
  \includegraphics[width=\columnwidth]{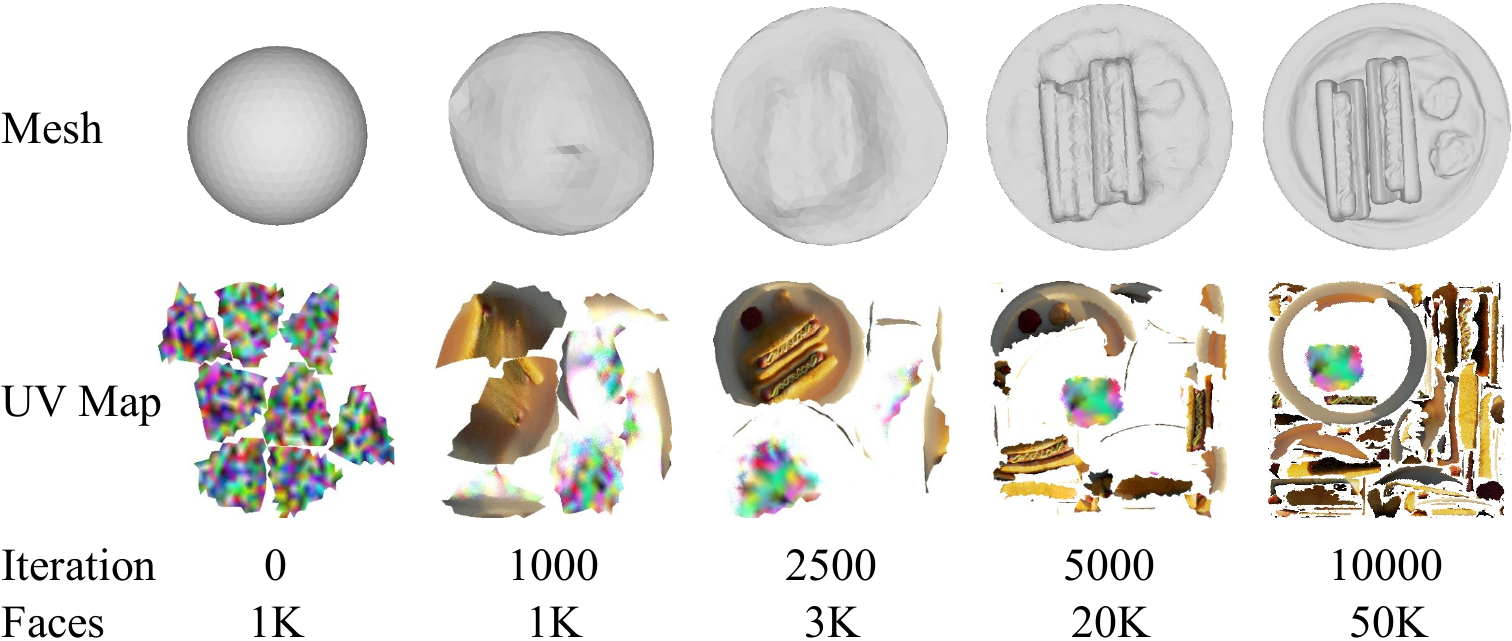}
  \caption{\textcolor{black}{Initialization experiment starting from a sphere with random colors.}}
  \label{fig:ablation-wo-init}
\end{figure}

\begin{figure}[t]
  \Description{Ablation results comparing stage-I optimization with and without periodic UV map reconstruction.}
  \centering
  \includegraphics[width=0.95\columnwidth]{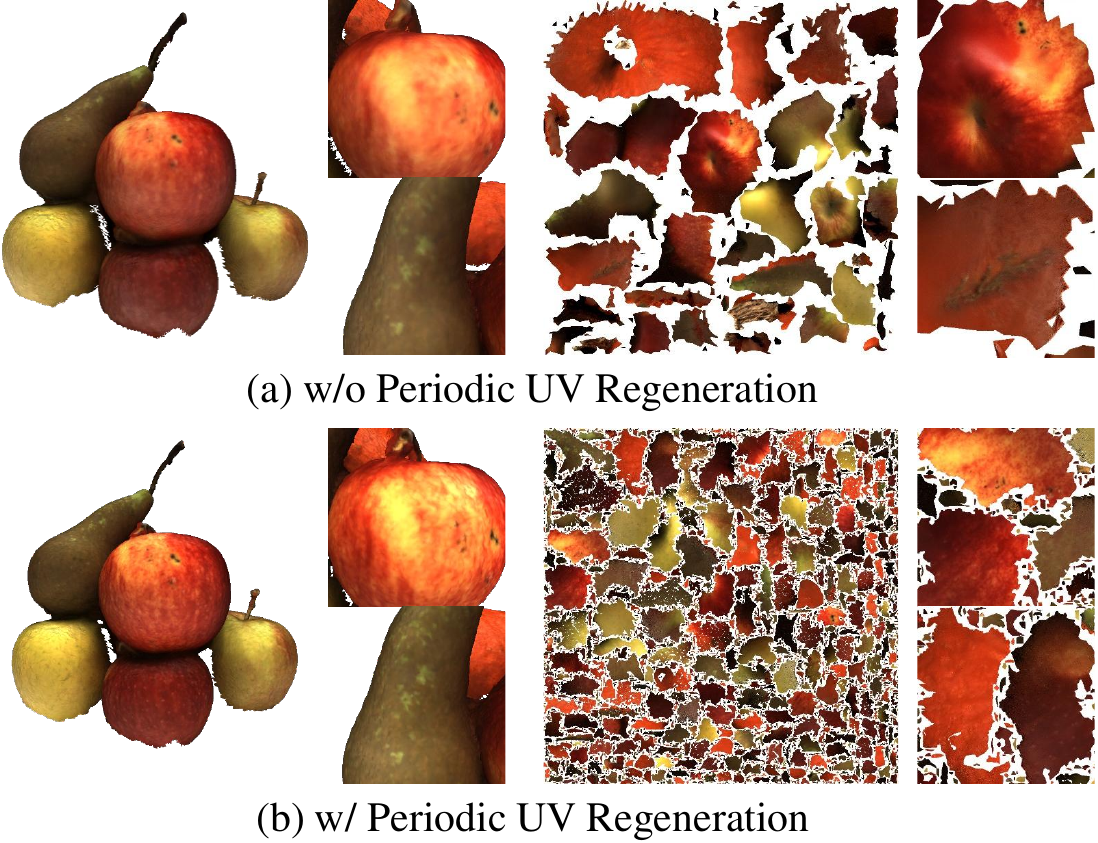}
  \caption{\textcolor{black}{Ablation of periodic UV map reconstruction in Stage I, which improves UV layout quality and preserves finer texture details.}}
  \label{fig:ablation-recreate-uvmap}
\end{figure}

\textbf{UV Reconstruction.} Fig.~\ref{fig:ablation-recreate-uvmap} compares the results with and without periodic UV map reconstruction in Stage I. The local updates in Sec.~\ref{sec:uv-maintenance} keep the UV mapping valid as the mesh topology changes, but repeated splitting and merging may gradually introduce atlas distortion, fragmented UV islands, and uneven texel allocation. Periodically reconstructing the UV atlas redistributes the texture space and provides more balanced resolution across the surface. As shown in Fig.~\ref{fig:ablation-recreate-uvmap}, this produces clearer local details and a more regular UV layout across the surface.
\end{revisedblock}

\begin{figure}[t]
  \Description{Ablation results for the second-stage PBR decomposition on Synthetic4Relight.}
  \centering
  \includegraphics[width=\columnwidth]{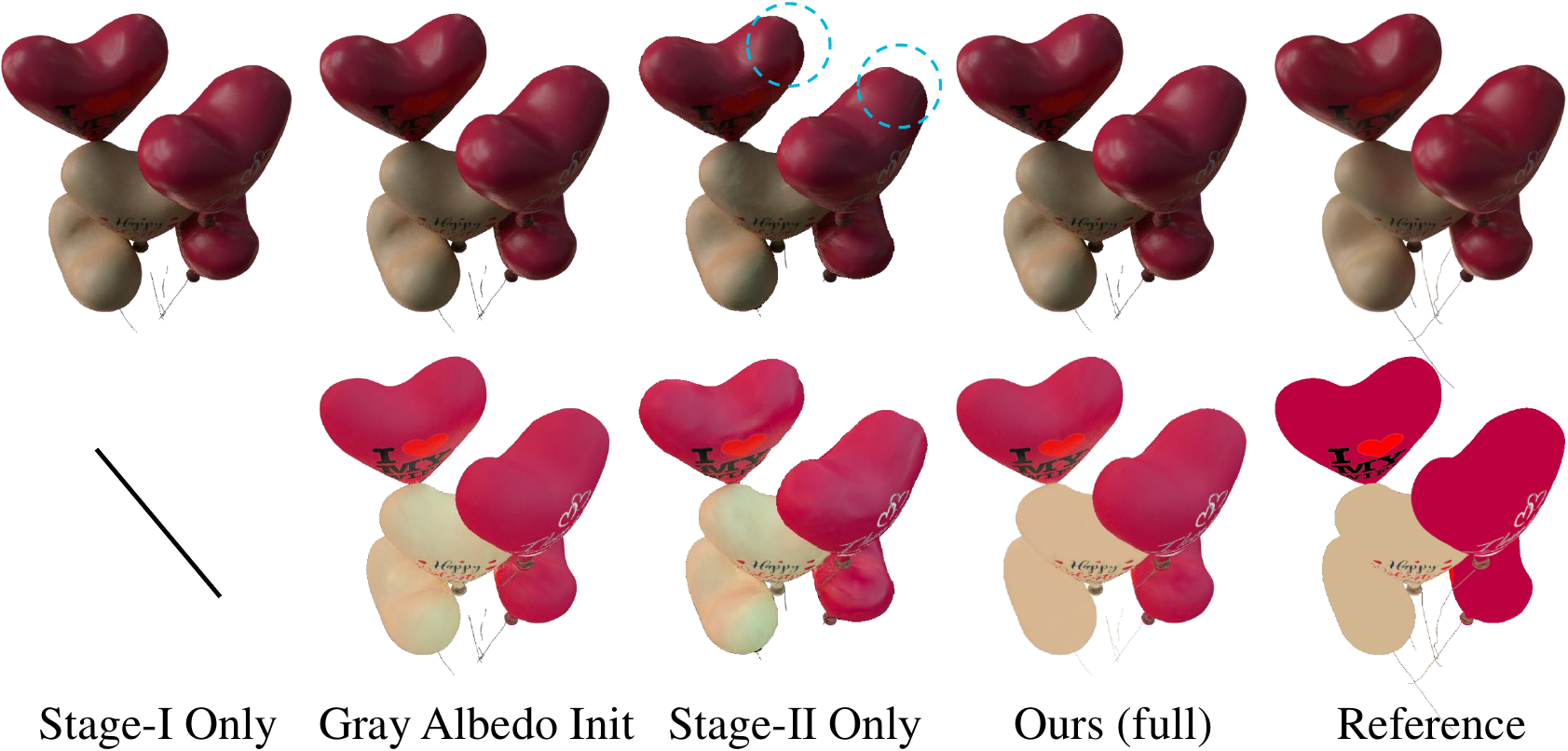}
  \caption{\added{Qualitative ablation of the second-stage PBR decomposition under different initialization and training settings. The top and bottom rows show relighting and recovered albedo, respectively.}}
  \label{fig:ablation-pbr}
\end{figure}

\begin{figure}[t]
  \Description{Qualitative ablation of Stage-II regularization terms, showing recovered albedo, roughness, and environment maps.}
  \centering
  \includegraphics[width=0.95\columnwidth]{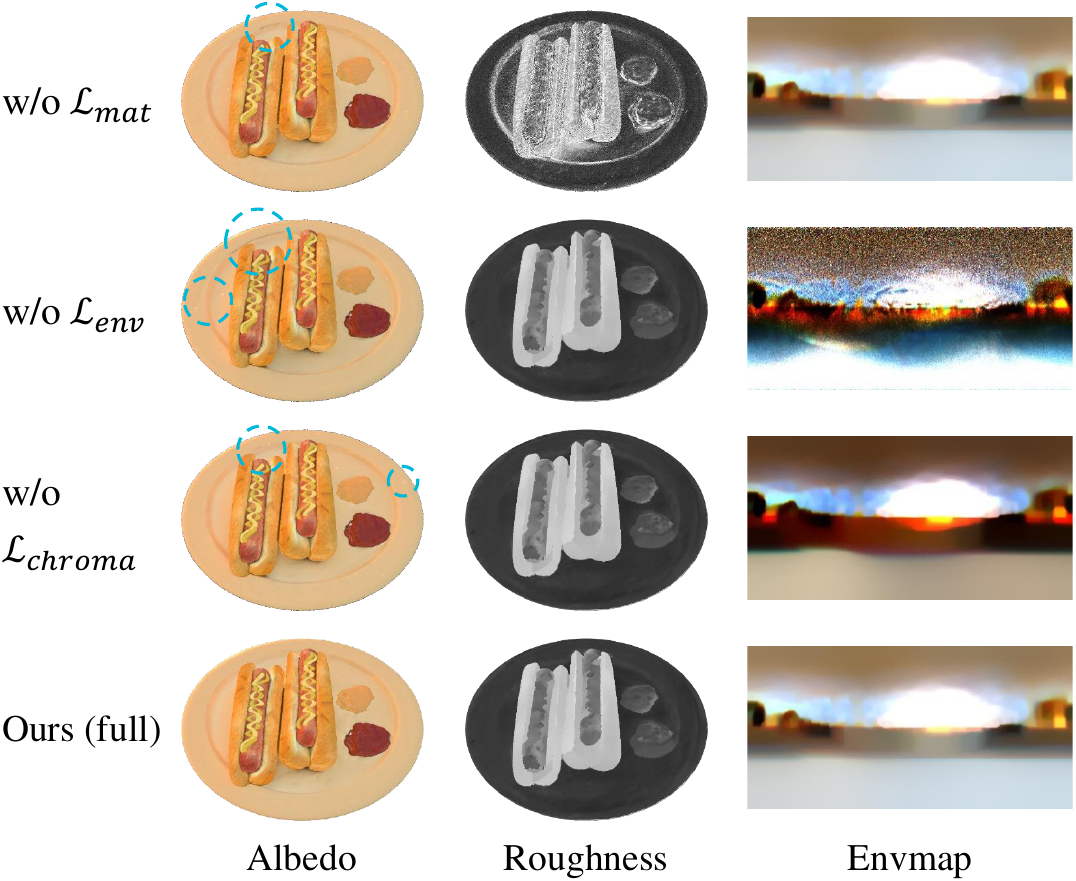}
  \caption{\added{Qualitative ablation of the Stage-II regularization terms. Cyan circles highlight representative material artifacts.}}
  \label{fig:pbr_reg_ablation}
\end{figure}

\begin{addedblock}
\textbf{PBR Stage.} We analyze the effect of the second-stage PBR decomposition on Synthetic4Relight, with results shown in Table~\ref{tab:ablation}(b) and Fig.~\ref{fig:ablation-pbr}. Stage-I Only performs only the first-stage training. It can fit the input illumination well, but its RGB texture still mixes intrinsic color, shadows, and highlights, making it unsuitable for relighting. Gray Base-Color Init starts Stage II from a gray base-color map instead of the RGB texture learned in Stage I. It recovers the main colors and lighting changes after PBR optimization, but produces noisier recovered albedo maps. Stage-II Only skips the first-stage mesh-UV carrier reconstruction. As a result, the mesh remains rough and does not fit the object geometry well, leading to local appearance errors on the balloon surfaces.

\textbf{Indirect Lighting.} Finally, we analyze the role of one-bounce indirect lighting during training and evaluation. As shown in Table~\ref{tab:ablation}(b), Direct Only gives lower relighting PSNR than settings with indirect lighting. Eval-only Indirect adds the one-bounce term only during evaluation and improves over Direct Only, but remains below the full method. No Secondary Material uses indirect lighting during both training and evaluation, but it does not query material information at secondary hit points. Its relighting metrics are close to the full method, while its albedo metrics are still lower. By querying shared UV-PBR materials at secondary hit points, the full setting achieves the best overall results.

\textbf{Stage-II Regularization.} We further ablate the three regularization terms in the Stage-II objective. As shown in Table~\ref{tab:pbr_reg_ablation} and Fig.~\ref{fig:pbr_reg_ablation}, removing $\mathcal{L}_{\mathrm{mat}}$ reduces both relighting and material-recovery quality, and produces strong high-frequency noise in the roughness map. Without $\mathcal{L}_{\mathrm{env}}$, the recovered environment map becomes noticeably noisy, which also degrades albedo recovery and relighting performance. Removing $\mathcal{L}_{\mathrm{chroma}}$ has a smaller quantitative effect, but causes object colors to leak into the recovered environment map and introduces local color artifacts in the albedo.

\textbf{Hyperparameter Sensitivity.} We vary five Stage-II hyperparameters on Synthetic4Relight while keeping the others fixed. As shown in Table~\ref{tab:hyperparameter_sensitivity}, Relight PSNR and Albedo PSNR vary by at most 0.15 dB and 0.11 dB, respectively. This indicates that ExMesh++ is not sensitive to moderate changes in these hyperparameters.
\end{addedblock}

\begin{addedblock}
\section{Limitations}
\label{sec:limitation}
The current appearance model uses an opaque, isotropic metallic-roughness BRDF. It does not represent more complex material effects such as anisotropic reflection, transmission, or subsurface scattering. In benchmark experiments, we disable metallic optimization to ensure a consistent comparison. The evaluation datasets do not provide ground-truth metallic maps, and several compared methods do not estimate a metallic channel. The optional metallic map is therefore demonstrated through downstream editing, but its recovery accuracy is not quantitatively evaluated.

\begin{table}[t]
  \caption{\added{Ablation of Stage-II regularization terms on Synthetic4Relight.}}
  \label{tab:pbr_reg_ablation}
  \centering
  \footnotesize
  \setlength{\tabcolsep}{3pt}
  \renewcommand{\arraystretch}{1.1}
  \resizebox{0.9\columnwidth}{!}{%
  \begin{tabular}{lccccc}
    \toprule
    \multirow{2}{*}{Setting}
    & \multicolumn{2}{c}{Relighting}
    & \multicolumn{2}{c}{Albedo}
    & \multicolumn{1}{c}{Roughness} \\
    & PSNR$\uparrow$ & LPIPS$\downarrow$
    & PSNR$\uparrow$ & LPIPS$\downarrow$
    & MSE$\downarrow$ \\
    \midrule
    w/o $\mathcal{L}_{\mathrm{mat}}$ & 33.76 & 0.066 & 29.81 & 0.077 & 0.012 \\
    w/o $\mathcal{L}_{\mathrm{env}}$ & 33.96 & 0.062 & 29.65 & 0.072 & 0.005 \\
    w/o $\mathcal{L}_{\mathrm{chroma}}$ & 34.01 & 0.063 & 29.84 & 0.068 & 0.005 \\
    Ours (full) & 34.19 & 0.061 & 30.16 & 0.068 & 0.005 \\
    \bottomrule
  \end{tabular}}
\end{table}

\begin{table}[t]
    \centering
    \caption{\added{Stage-II hyperparameter sensitivity on Synthetic4Relight.}}
    \label{tab:hyperparameter_sensitivity}
    \setlength{\tabcolsep}{4.5pt}
    \begin{tabular}{lccc}
        \toprule
        Parameter &
        Tested values &
        Relight PSNR &
        Albedo PSNR \\
        \midrule
        $\lambda_{\mathrm{mat}}$
        & $\{0.05,\,0.1,\,0.2\}$
        & 34.09--34.19
        & 30.10--30.19 \\

        $\lambda_{\mathrm{env}}$
        & $\{0.05,\,0.1,\,0.2\}$
        & 34.11--34.21
        & 30.08--30.16 \\

        $\lambda_{\mathrm{chroma}}^{0}$
        & $\{0.05,\,0.1,\,0.2\}$
        & 34.15--34.20
        & 30.13--30.18 \\

        $\lambda_{\mathrm{ind}}^{\max}$
        & $\{0.5,\,1.0,\,1.5\}$
        & 34.13--34.19
        & 30.05--30.16 \\

        $N_b$
        & $\{8,\,16,\,32\}$
        & 34.08--34.23
        & 30.03--30.19 \\
        \bottomrule
    \end{tabular}
\end{table}

Our current light-transport model considers only one-bounce diffuse interreflection. It captures local color bleeding and illumination in occluded regions, but does not model multi-bounce transport or indirect specular reflection. Extending the renderer to richer light transport could improve physical realism, although it would increase optimization and rendering costs.

Finally, periodic UV regeneration currently relies on the CPU-based xatlas implementation, which becomes a computational bottleneck for high-resolution meshes. Our current experiments focus on object-level assets. Scaling the pipeline to larger scenes with many components, large texture atlases, and complex meshes requires more efficient UV parameterization and memory management.

\section{Conclusion}
\label{sec:conclusion}

We presented ExMesh++, a staged framework for reconstructing editable and relightable UV-PBR mesh assets from multi-view images. Stage I directly refines explicit mesh geometry and topology through vertex splitting and merging, while maintaining a valid UV parameterization. Stage II freezes the reconstructed mesh-UV carrier and decomposes its appearance into UV-space PBR material maps and environment lighting. The shared UV-PBR representation further supports one-bounce diffuse indirect illumination without an additional learned residual appearance field. Experiments on synthetic and real-captured datasets demonstrate competitive geometry, material-recovery, and relighting performance. The exported assets can also be directly edited, relit, and composed with artist-created content in standard DCC workflows.

\end{addedblock}

\bibliographystyle{ACM-Reference-Format}
\bibliography{acmart}

\end{document}